\documentclass[10pt, aps, prl, twocolumn, superscriptaddress, nofootinbib]{revtex4-2}

\usepackage{amssymb,amsmath,mathrsfs,enumerate}
\usepackage{graphicx}
\usepackage{feynmp-auto}
\usepackage{mathtools}
\usepackage{caption}
\usepackage{subcaption}
\usepackage[dvipsnames]{xcolor}
\usepackage[colorlinks=true, linkcolor=blue, urlcolor=Blue, citecolor=Mulberry]{hyperref}
\usepackage{orcidlink}
\usepackage{multirow}
\usepackage{booktabs}
\usepackage[normalem]{ulem}
\usepackage{microtype}  % nicer spacing (optional)
\usepackage{ragged2e}   % for \plotalign

\newcommand{\blfootnote}[1]{%
  \begingroup
  \renewcommand{\thefootnote}{}\footnote{#1}%
  \addtocounter{footnote}{-1}%
  \endgroup
}

\makeatletter
\patchcmd{\@makecaption}{\ignorespaces}{\justifying\ignorespaces}{}{}
\makeatother

\begin{document}

\title{Dark Recipe for the First Giants: From Population III Stars to Early Supermassive Black Holes via Dark Matter Capture}

\author{Sulagna Bhattacharya \orcidlink{0000-0002-4034-177X}}
\email{sulagna@theory.tifr.res.in}
\affiliation{Tata Institute of Fundamental Research, Homi Bhabha Road, Mumbai 400005, India}

\author{Debajit Bose \orcidlink{0000-0001-8594-8885}}
\email{debajitbose550@gmail.com}
\affiliation{Centre for High Energy Physics, Indian Institute of Science, C. V. Raman Avenue, Bengaluru 560012, India}

\author{Basudeb Dasgupta \orcidlink{0000-0001-6714-0014}}
\email{bdasgupta@theory.tifr.res.in}
\affiliation{Tata Institute of Fundamental Research, Homi Bhabha Road, Mumbai 400005, India}

\author{\\Jaya Doliya \orcidlink{0009-0003-8407-7442}}
\email{jayadoliya@iisc.ac.in}
\affiliation{Centre for High Energy Physics, Indian Institute of Science, C. V. Raman Avenue, Bengaluru 560012, India}

\author{Ranjan Laha \orcidlink{0000-0001-7104-5730}}
\email{ranjanlaha@iisc.ac.in}
\affiliation{Centre for High Energy Physics, Indian Institute of Science, C. V. Raman Avenue, Bengaluru 560012, India}

\date{\today}
%
%
%%%%%%%%%%%%%%%%%%%%%%%%%%%%%%%%%%%%%%%%%%%%%%%%%%%%%%%%%%%%%%%%%%%%%%%%%%%
%%%%%%%%%%%%%%%%%%%%%%%%       ABSTRACT       %%%%%%%%%%%%%%%%%%%%%%%%%%%%%
\begin{abstract}

The presence of supermassive black holes (SMBHs) at high redshifts ($z>5$), as revealed by James Webb Space Telescope (JWST), challenges standard black hole (BH) formation scenarios. We propose a mechanism in which non-annihilating dark matter (DM) with non-gravitational interactions with the Standard Model (SM) particles accumulates inside Population III (Pop III) stars, inducing their premature collapse into BH seeds having the same mass as the parent star. Owing to their early formation, these seeds can accrete for longer periods and grow into the SMBHs observed at early cosmic times. Focusing on spin-dependent (SD) DM–proton interactions, we identify regions of parameter space that account for the observed high-redshift SMBH population, their mass function, and the SMBH–stellar mass relation. Portions of this parameter space are testable by forthcoming direct detection experiments. The scenario may lead to distinctive gravitational wave (GW) signatures from SMBH mergers, accessible to Laser Interferometer Space Antenna (LISA) and pulsar timing array (PTA) observations.

\end{abstract}
%%%%%%%%%%%%%%%%%%%%%%%%%%%%%%%%%%%%%%%%%%%%%%%%%%%%%%%%%%%%%%%%%%%%%%%%%%%
%%%%%%%%%%%%%%%%%%%%%%%%%%%%%%%%%%%%%%%%%%%%%%%%%%%%%%%%%%%%%%%%%%%%%%%%%%%
%
%
\maketitle
\blfootnote{S.B., D.B., and J.D. contributed equally to this work.}

%
%
%%%%%%%%%%%%%%%%%%%%%%%%%%%%%%%%%%%%%%%%%%%%%%%%%%%%%%%%%%%%%%%%%%%%%%%%%%%
%%%%%%%%%%%%%%%%%%%%%%%%       INTRO      %%%%%%%%%%%%%%%%%%%%%%%%%%%%%%%%%
\section{Introduction}
\label{sec:intro}

Understanding the origin of SMBHs in the early Universe remains a central open problem in cosmology. Optical and infrared surveys, such as the Sloan Digital Sky Survey (SDSS)\,\cite{SDSS:2001emm, SDSS:2003iyw}, Canada-France High-redshift Quasar Survey (CFHQS)\,\cite{Willott:2007rm, Willott:2009wv}, Subaru High-z Exploration of Low-Luminosity Quasars (SHELLQs)\,\cite{matsuoka2018subaru}, and X-ray observatories like Chandra\,\cite{Bogdan:2023ilu} and eROSITA\,\cite{Medvedev:2020urn} have discovered SMBHs at redshifts $z \gtrsim 5$\,\cite{Inayoshi:2019fun}. Additionally, JWST is discovering SMBHs at high redshifts at a remarkable pace\,\cite{Maiolino:2023zdu, Bogdan:2023ilu, Kovacs:2024zfh, Ubler2025BlackTHUNDEREF}. JWST has also revealed a new population of red-tinted cosmological sources, known as Little Red Dots (LRDs)\,\cite{Labb2022APO, Matthee:2023utn, Kokorev2024ACO, Zhang2025JWSTII, Ronayne2025MEGASS, Tanaka2025DiscoveryOA, Maiolino:2025tih, Greene:2025njv, Fu2025DiscoveryOT}, that are believed to potentially host SMBHs at their centers that are in the early phases of active galactic nuclei (AGN) activity. The characteristic red-tint has been hypothesized to be attributed to the dense dust outflows from these young AGNs that enshroud the central region and significantly redden the emitted light\,\cite{Pacucci:2024tws, Graaff2024RUBIESAC, ronayne2025mega, Zhuang2025NEXUSAS, Quadri:2025zvj, Inayoshi:2025hdr, Juodvzbalis2025ADB, Kim2025TheBA, Delvecchio2025AGNheatedDR, Pacucci:2025ojp, Jones2025TheB, 2025arXiv251110725H, Graaff2025LittleRD, Inayoshi2025ACE}. Explaining the rapid assembly of such massive objects poses a key theoretical challenge.

Several pathways have been proposed for the formation of SMBHs at high redshifts.\,\textit{(i)}\,One of the astrophysical scenarios involves remnants of the first generation of stars, i.e., Pop III stars. These metal-free stars are expected to form within dark matter halos of mass $M_{\rm halo} \gtrsim 10^5 – 10^6 \, {\rm M_\odot}$ at $z \sim 15 - 30$, where baryonic matter can cool efficiently through molecular ${\rm H_2}$ transitions to form these stars\,\cite{Hirano:2013lba, Hirano:2015wxa, Klessen:2023qmc, Ho:2025cou, Ellis2025GalaxiesAB, Glover2025TheFS}. Heavier Pop III stars $(M_s \gtrsim 260 \, {\rm M_\odot})$ can collapse directly into BHs of similar mass, which can then accrete to form SMBHs\,\cite{Sanati2025TheEA}.\,\textit{(ii)}\,Metals expelled from Pop III stars enrich gas clouds, inducing fragmentation and formation of low-mass stars. These stars, formed in dense stellar clusters, undergo repeated collisions to form heavier stars that can collapse into BHs with masses $\mathcal{O}(10^2 - 10^3) \, {\rm M_\odot}$ at $z \sim 10 - 20$\,\cite{Omukai:2008wv, Bernadetta:2008bc, Rantala2025ARC, Reinoso2025MassiveBH, Bhowmick2025HeavySA, Dekel2025FromFS}.\,\textit{(iii)}\,Direct collapse of metal-free gas clouds in the early Universe can form massive BHs with masses $\mathcal{O}(10^4 - 10^6) \, {\rm M_\odot}$ at $z \sim 8 - 17$\,\cite{Eisenstein:1994nh, Ferrara:2014wua, Jeon:2024iml, Jeon2025TheEB, Kimura2025MassiveBH, Cenci2025LittleRD, Jeon2025LittleRD, Santarelli:2025uck}. Different beyond the Standard Model (BSM) frameworks have been proposed that can provide the additional Lyman–Werner (LW) photon background to accelerate the collapse of pristine gas clouds\,\cite{Friedlander:2022ovf, Lu:2023xoi, Lu:2024zwa, Jiao:2025kpn, Lu:2025kbe, Qin:2025ymc, Zwick:2025eik, Zhang:2025grn, Aggarwal:2025pit}. SMBH formation via self-interacting DM, primordial black holes, dark stars, and several other mechanisms have also been studied\,\cite{Balberg:2002ue, Pollack:2014rja, Lu:2024xnb, Roberts:2024wup, Ziparo:2024nwh, Buckley:2024eoe, Schwemberger:2024rvu, Jiang:2025jtr, Shen:2025evo, Roberts:2025poo, Li2025TheLittleDD, Tran:2025riw, Chen:2025jch, Feng:2025rzf, Kiyuna2025SuperEddingtonGC, Prole:2025snf, Dayal:2025aiv, Yajima2025FormationOO, Cyr:2025tkt, Mould2025IfQF, Zhang:2025tgm, Roberts2025LittleRD, Nandal:2025xvv, Imai:2025ate, Dong:2025iez, McDonald:2025rfm, Chen:2025bqb, Ellis:2025dpw, Kritos:2025aqo, Sanchis-Gual:2025lbp, Topalakis:2025gux, Chen:2025vga, Ilie:2025zzj, Freese:2025dmo, Pauchet2025DarkMatterPoweredPI}.

We explore the scenario where non-annihilating DM particles (e.g., asymmetric DM\,\cite{Petraki:2013wwa, Zurek:2013wia}) scatter non-gravitionally with Pop III star to get captured and accumulate sufficiently to trigger tiny BH formation inside the stellar core\,\cite{Ellis:2021ztw, Diks:2024cww}, which can subsequently accrete the stellar material to form seed of mass $M_{\rm seed} \sim \mathcal{O}(10-100) \, {\rm M_\odot}$. We first try to explain the observed SMBHs ($\gtrsim 10^6 \, M_\odot$) at $z_{\rm obs} \gtrsim 5$ along with the estimation of the host galaxy stellar mass for an allowed benchmark point of the DM parameter space. For the same benchmark point, we also calculate the SMBH mass function at observed redshifts. Furthermore, we study the associated GWs formed from these isolated SMBH mergers and the stochastic background and their detectability with LISA\,\cite{2017arXiv170200786A} and PTAs\,\cite{NANOGrav:2023ctt}.
\begin{figure}[t]
    \centering
    \includegraphics[width=1\linewidth]{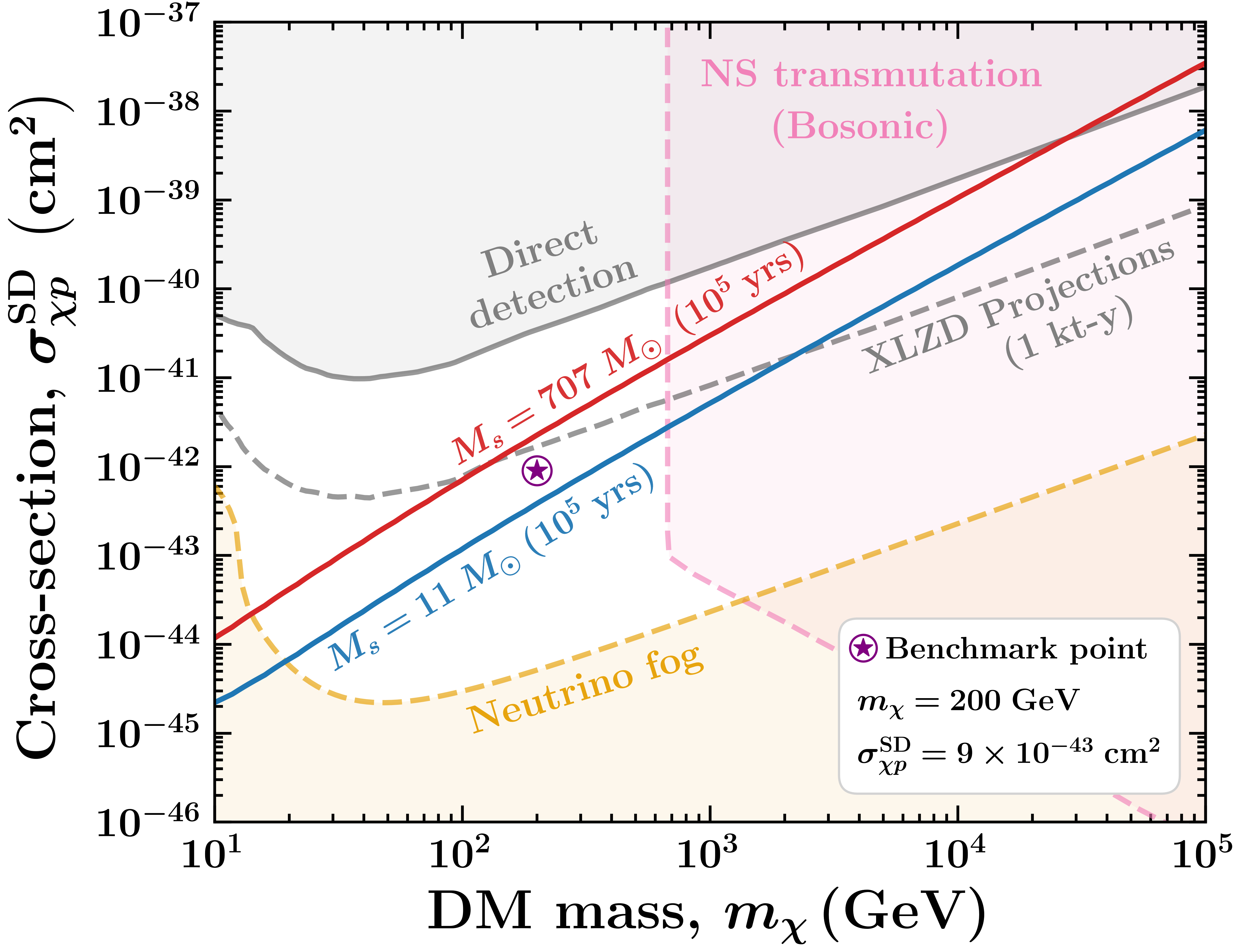}
    \caption{\justifying Contours of seed BH formation time of $10^5 \, {\rm yrs}$ for the lowest (blue) and the highest (red) mass Pop III stars for DM–proton SD interaction ($\sigma^{\rm SD}_{\chi p}$) are shown along with competitive limits in the parameter space. These two masses represent the lowest and highest masses across all redshifts from\,\cite{Hirano:2015wxa}. The gray, pink, and orange shaded regions represent the exclusion limits from combined PICO-60\,\cite{PICO:2019vsc} and LZ\,\cite{LZ:2024zvo}, bounds from neutron star transmutation\,\cite{Garani:2018kkd}, and the neutrino fog\,\cite{OHare:2021utq}, respectively. The purple point denotes the chosen DM benchmark point. The gray dashed curve represents the projected $1 \,$ktonne-year (kt-y) SD sensitivity of XLZD \cite{XLZD:2024nsu}, obtained by scaling the current SD LZ limits.}
    \label{fig:bhform_isochrone}
\end{figure}
%
%
%%%%%%%%%%%%%%%%%%%%%%%%%%%%%%%%%%%%%%%%%%%%%%%%%%%%%%%%%%%%%%%%%%%%%%%%%%%
%%%%%%%%%%%%%%%%%%%%%%%%%%%%%%%%%%%%%%%%%%%%%%%%%%%%%%%%%%%%%%%%%%%%%%%%%%%

%
%
%%%%%%%%%%%%%%%%%%%%%%%%%%%%%%%%%%%%%%%%%%%%%%%%%%%%%%%%%%%%%%%%%%%%%%%%%%%
%%%%%%%%%%%%%%%%%%%       BH FORMATION TIMESCALES     %%%%%%%%%%%%%%%%%%%%%
\section{SMBH Formation}
\label{bh_form_time}

DM particles $(\chi)$ can scatter with SM constituents of Pop III stars and lose sufficient energy to get captured without being evaporated away, provided their mass $m_\chi \gtrsim 10 \, {\rm GeV}$\,\cite{Garani:2021feo, Ellis:2021ztw}. We consider the optically thin regime, which makes the single-scatter approximation valid. The total timescale to form the SMBH seed through DM capture is
\begin{equation}\label{eq:t_bhform}
    \tau = \tau_{\rm accum}  + \tau_{\rm therm} +  \tau_{\rm col} + \tau_{\rm Bondi},
\end{equation}
where $\tau_{\rm accum}$ is the accumulation timescale for DM to trigger the collapse, which is determined by $m_\chi C_\chi \tau_{\rm accum} = \max[M_{\rm self}, M_{\rm Ch}]$, with $C_\chi$ being the DM capture rate. Here, $M_{\rm self}$ is the self-gravitating mass and $M_{\rm Ch}$ is the Chandrasekhar mass arising from Fermi degeneracy pressure. For bosonic DM, the balancing quantum pressure comes from the zero-point energy of the quantum states or from self-interaction (if it exists)\,\cite{Chavanis:2011cz, Bell:2013xk, Eby:2015hsq}. For the considered DM masses, $M_{\rm self}$ overwhelms the degeneracy and quantum pressure of fermionic and bosonic DM, respectively, which makes $\tau_{\rm accum}$ and, consequently, the whole framework applicable to both. The thermalization time is denoted by $\tau_{\rm therm}$. The timescale over which the DM core contracts down to the Schwarzschild radius $(r_{\rm Sch})$ is denoted by $\tau_{\rm col}$, and $\tau_{\rm Bondi}$ is the time required for the tiny BH to consume the parent star via Bondi accretion\,\cite{Bondi:1952ni}. Note that, for the parameter space of our interest, the seed formation timescale is dominated by the thermalization timescale. Therefore, the nature of accretion for stellar consumption does not matter much. For our chosen DM parameters, the tiny BH evaporation is negligible\,\cite{Ellis:2021ztw}. To calculate the capture rate and the associated timescales, we use the properties of Pop III stars obtained from \texttt{MESA}\,\cite{Paxton:2010ji, Paxton2013MODULESFE, Paxton2015MODULESFE, Paxton:2017eie, Paxton2019ModulesFE, MESA:2022zpy}. All calculational details are outlined in the \emph{Supplemental Material}.
%
%
%%%%%%%%%%%%%%%%%%%%%%%%%%%%%%%%%%%%%%%%%%%%%%%%%%%%%%%%%%%%%%%%%%%%%%%%%%%
\begin{figure*}[t]
    \centering
    \begin{subfigure}{0.495\textwidth}
        \centering
        \includegraphics[width=0.975\linewidth]{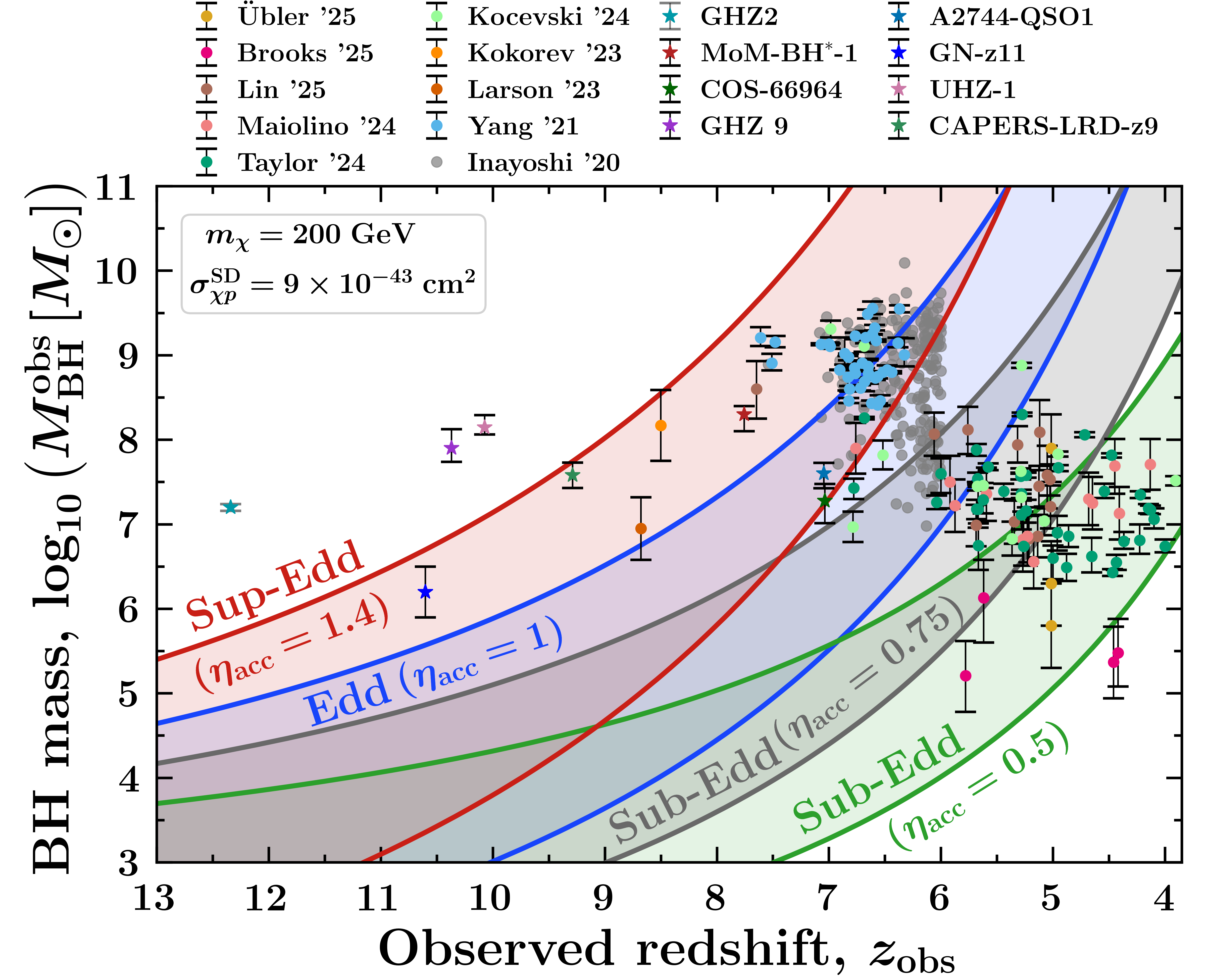}
        \caption{}
        \label{sf:SMBH_mass_redshift}
    \end{subfigure}
    \begin{subfigure}{0.495\textwidth}
        \centering
        \includegraphics[width=0.975\linewidth]{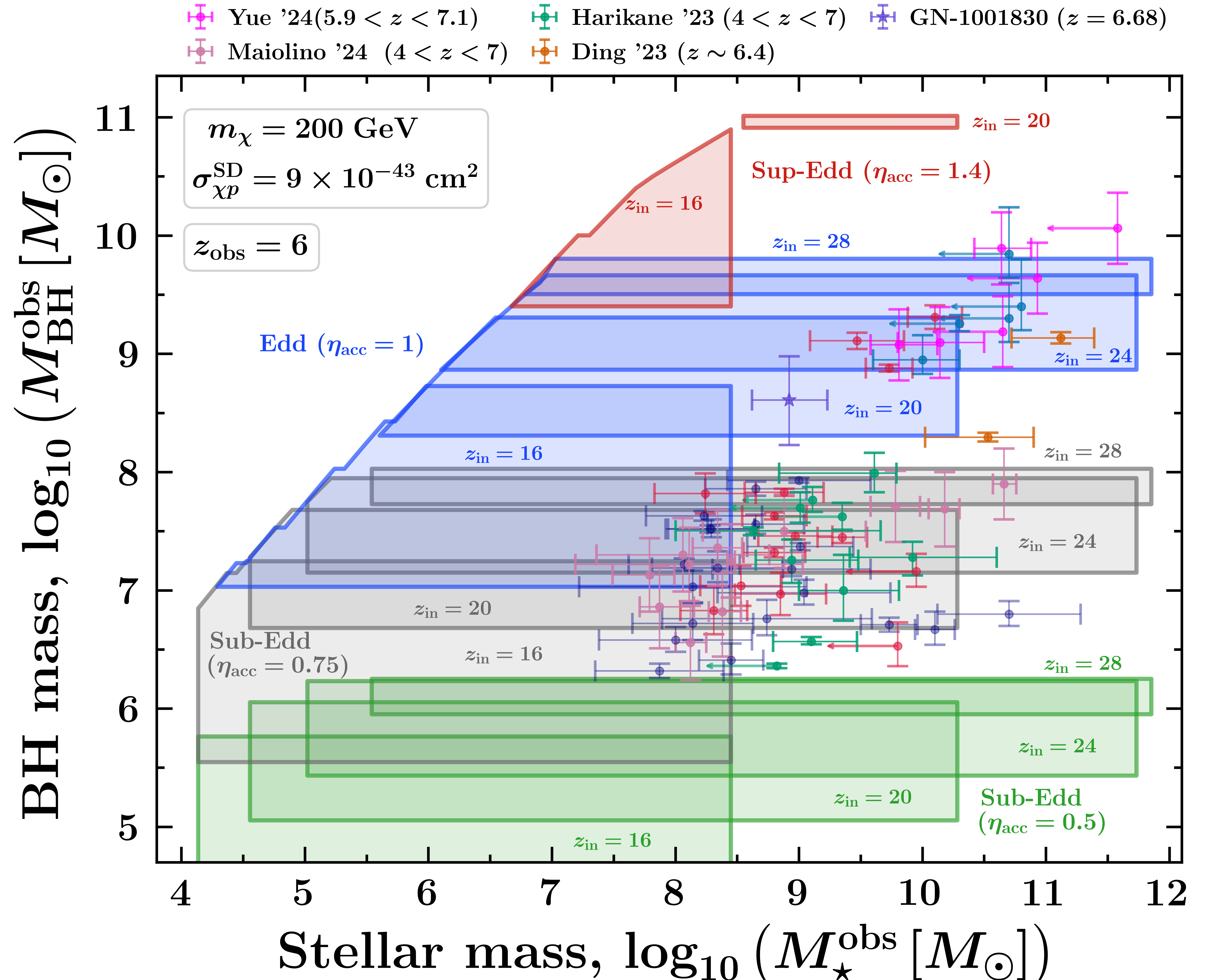}
        \caption{}
        \label{sf:MBH_vs_Mstar}
    \end{subfigure}
    \caption{\justifying \textbf{Left:} SMBH masses as a function of observed redshifts for our chosen DM benchmark point, shown for Eddington (blue: $\eta_{\rm acc}=1$), sub-Eddington (green: $\eta_{\rm acc}= 0.5$ and gray: $\eta_{\rm acc}= 0.75$), and super-Eddington (red: $\eta_{\rm acc}=1.4$) accretion rates, along with observed SMBHs\,\cite{Inayoshi:2019fun, Kovacs:2024zfh, Maiolino:2023zdu, Taylor2025CAPERSLRDz9AG, Yang:2021imt, Goulding:2023gqa, Larson2023ACD, Kokorev2023UNCOVERAN, Maiolino:2023bpi, Furtak2023AHB, Naidu:2025rpo, Lin2025BridgingQA, Akins2024StrongRE}. \textbf{Right:} SMBH mass and host galaxy stellar mass relation at $z_{\rm obs} = 6$ is shown for Eddington (blue, $\eta_{\rm acc}=1$), sub-Eddington (green: $\eta_{\rm acc}= 0.5$ and gray: $\eta_{\rm acc}= 0.75$), and super-Eddington (red: $\eta_{\rm acc}=1.4$) accretion rates, together with the observed SMBH and stellar mass data at similar redshifts\,\cite{Stone2023UndermassiveHG, Yue2023EIGERVC, Maiolino:2023bpi, Harikane2023AJF, Ding2022DetectionOS, Juodbalis2024ADO}. }
    \label{fig:SMBH_mass_redshift_MBH_Mstar_combined}
\end{figure*}
%%%%%%%%%%%%%%%%%%%%%%%%%%%%%%%%%%%%%%%%%%%%%%%%%%%%%%%%%%%%%%%%%%%%%%%%%%%
%
%

In~Fig.\,\ref{fig:bhform_isochrone}, we show the isochrones, contours of the same BH-seed formation time, of $\tau = 10^5 \, {\rm yrs}$ $\left( \ll \tau^{\rm Pop III}_{\rm lifetime} \right)$ for the lowest $\left(11 \, M_\odot, \, 1.5 \, R_\odot, \, \tau^{\rm Pop III}_{\rm lifetime} = 14.5 \, {\rm Myr} \right)$ and the highest Pop III masses $\left(707 \, M_\odot, \, 16.4 \, R_\odot, \, \tau^{\rm Pop III}_{\rm lifetime} = 1.9 \, {\rm Myr} \right)$ from \cite{Hirano:2015wxa} for SD DM–proton scattering. The purple point corresponds to the considered benchmark parameters ($m_\chi = 200 \, {\rm GeV}, \, \sigma_{\chi p}^{\rm SD} = 9 \times 10^{-43} \, {\rm cm^2}$) that evade existing constraints. Although we present only one benchmark point, Fig.~\ref{fig:bhform_isochrone} shows that a broader parameter space, including some regions within the neutrino fog, leads to similar $\tau$. Some portions of the allowed DM parameter space can be probed by the XLZD experiment\,\cite{XLZD:2024nsu}. We also explore whether DM–electron scattering can explain these SMBH observations (see the \emph{Supplemental Material}).

For a Pop III star forming at $z_{\rm in}(t_{\rm in})$, the seed BH forms at $z_{\rm seed}(t_{\rm seed} = t_{\rm in} + \tau)$. The seed BH grows via accretion, as defined in Eq.~\eqref{eq:MBH_obs}. We consider three growth rates: Eddington, sub-Eddington, and super-Eddington. The corresponding Salpeter time is $t_{\rm Sal} \approx \left( \dfrac{\epsilon_r}{0.1} \right) \, 45.1 \, {\rm Myr}$ \cite{Salpeter1964AccretionOI}, where the radiative efficiency $\epsilon_r$ is taken to be $0.1$ that depends on the BH angular momentum\,\cite{Shapiro:2004ud}. Our choice of Bondi accretion for stellar consumption by the tiny BH is conservative; however, the subsequent (super-/sub-) Eddington accretion rate for seed BH growth is demanded by the requirement to explain the observational data. With the Bondi accretion rate, a $100 \, M_{\odot}$ seed will need a boost factor of $\sim 10^4$ in the ambient halo matter density. The observed BH mass at an observed redshift, $z_{\rm obs}(t_{\rm obs})$, is
\begin{equation}\label{eq:MBH_obs}
M_{\rm BH}^{\rm obs} = M_{\rm Seed} \times \exp \left[ \dfrac{\eta_{\rm acc} (1 - \epsilon_r) (t_{\rm obs} - t_{\rm seed})}{t_{\rm Sal}} \right],
\end{equation}
where $\eta_{\rm acc}$ denotes the accretion efficiency, taken to be $1.0$ for Eddington accretion, $0.5$ and $0.75$ for sub-Eddington accretion, and $1.4$ for super-Eddington growth. In Fig.~\ref{sf:SMBH_mass_redshift}, we show the observed SMBH masses predicted in our framework as a function of $z_{\rm obs}$. Each band corresponds to a different accretion efficiency obtained for four initial Pop III formation redshifts, $z_{\rm in}$, (16, 20, 24, and 28)\,\cite{Hirano:2015wxa}. The upper and lower edges of each band correspond to the heaviest Pop III star at $z=28$ and the lightest Pop III star at $z=16$, respectively. For comparison, we overlay the observed SMBHs\,\cite{Inayoshi:2019fun, Kovacs:2024zfh, Maiolino:2023zdu, Taylor2025CAPERSLRDz9AG, Yang:2021imt, Goulding:2023gqa, Larson2023ACD, Kokorev2023UNCOVERAN, Maiolino:2023bpi, Furtak2023AHB, Naidu:2025rpo, Lin2025BridgingQA, Akins2024StrongRE}. Eddington and sub-Eddington accretion together can explain $\sim 81\%$ of the observed data, which increases to $\sim 99\%$ when super-Eddington accretion is included.
%
%
%%%%%%%%%%%%%%%%%%%%%%%%%%%%%%%%%%%%%%%%%%%%%%%%%%%%%%%%%%%%%%%%%%%%%%%%%%%
%%%%%%%%%%%%%%%%%%%%%%%%%%%%%%%%%%%%%%%%%%%%%%%%%%%%%%%%%%%%%%%%%%%%%%%%%%%

%%%%%%%%%%%%%%%%%%%%%%%%%%%%%%%%%%%%%%%%%%%%%%%%%%%%%%%%%%%%%%%%%%%%%%%%%%%
%%%%%%%%%%%%%%%%%%%%%       SMBH MASS FUNCTION       %%%%%%%%%%%%%%%%%%%%%%
\section{SMBH Mass Function}
\label{sec:smbh_mass_func}

The mechanism of DM capture within Pop III stars offers a plausible explanation for the observed SMBHs. We evaluate the SMBH mass function $(\Phi_{\rm BH}^{\rm obs})$ by modeling the DM halos and the Pop III star distribution at high redshifts. We consider halo masses spanning $M_{\rm halo} \in \left[ 10^6, \, 10^{9} \right] \, M_\odot$ to ensure the formation of Pop III stars within these halos. The density profile of DM is considered to follow the generalized Navarro–Frenk–White (gNFW) profile\,\cite{Navarro:1995iw, Navarro:1996gj, Klypin:2000hk} with an inner slope of $3/2$\,\cite{Moore:1999gc}. We assume a conservative value for the concentration parameter $c_{200} = 2$ based on Fig.~7 of~\cite{Correa:2015dva}, which shows the dependence of $c_{200}$ on halo masses up to $z = 20$, with typical values of $c_{200} \gtrsim 3$.

We assume one Pop III star at the central region of each halo (at $r_{\rm dis}$), where the DM density is $10^{11} \, {\rm GeV/cm^3}$. The mass of the star is drawn from\,\cite{Hirano:2015wxa}. Throughout the lifetime of the Pop III star, we assume that it does not migrate too much from this location. At $r_{\rm dis}$, the rotational and escape velocities of the halos are calculated to evaluate the DM capture rate (see Section~\ref{sec:dm_halo_details}). For a given halo mass at $z_{\rm in}$, we compute $\tau$ with the chosen benchmark parameters for each bin of the IMF constructed from\,\cite{Hirano:2015wxa}, followed by four accretion scenarios of the seed BH. With the fitting relation given in\,\cite{Fakhouri:2010st}, we calculate the evolved halo mass at $z_{\rm obs}$ and impose $M_{\rm BH}^{\rm obs} \leq M_{\rm halo}^{\rm obs}$. This process is carried out for $M_{\rm halo} \in \left[10^6, 10^{9}\right ]\, {\rm M_\odot}$ for the four chosen $z_{\rm in}$ and then summed at $z_{\rm obs}$. After tracing the evolution of halo masses up to $z_{\rm obs}$, the stellar mass of the halos $(M_{\star}^{\rm obs})$ is obtained by multiplying the DM halo mass by the stellar-to-halo mass fraction derived from the fitting of the \texttt{Data Release\,1} of the \texttt{UniverseMachine} code\,\cite{Behroozi2019_UM_DR1}. For $M_{\rm halo} < 10^{10.5} \, M_\odot$, where the fractions are unavailable, we adopted the lowest quoted value from the fitting. We plot the obtained stellar mass versus SMBH mass in Fig.~\ref{sf:MBH_vs_Mstar} at $z_{\rm obs} = 6$, along with the observations at similar redshifts\,\cite{Stone2023UndermassiveHG, Yue2023EIGERVC, Maiolino:2023bpi, Harikane2023AJF, Ding2022DetectionOS, Juodbalis2024ADO}. Collectively, Eddington and sub-Eddington accretion formalisms explain most of the observed data. Increasing observational exploration of high redshifts galaxies such as Abell~2744-QSO1 (at $z_{\rm obs} = 7$), which exhibit a high SMBH-to-galaxy stellar mass ratio, can provide a unique signature of our scenario.

\begin{figure}[t]
    \centering
    \includegraphics[width=0.975\linewidth]{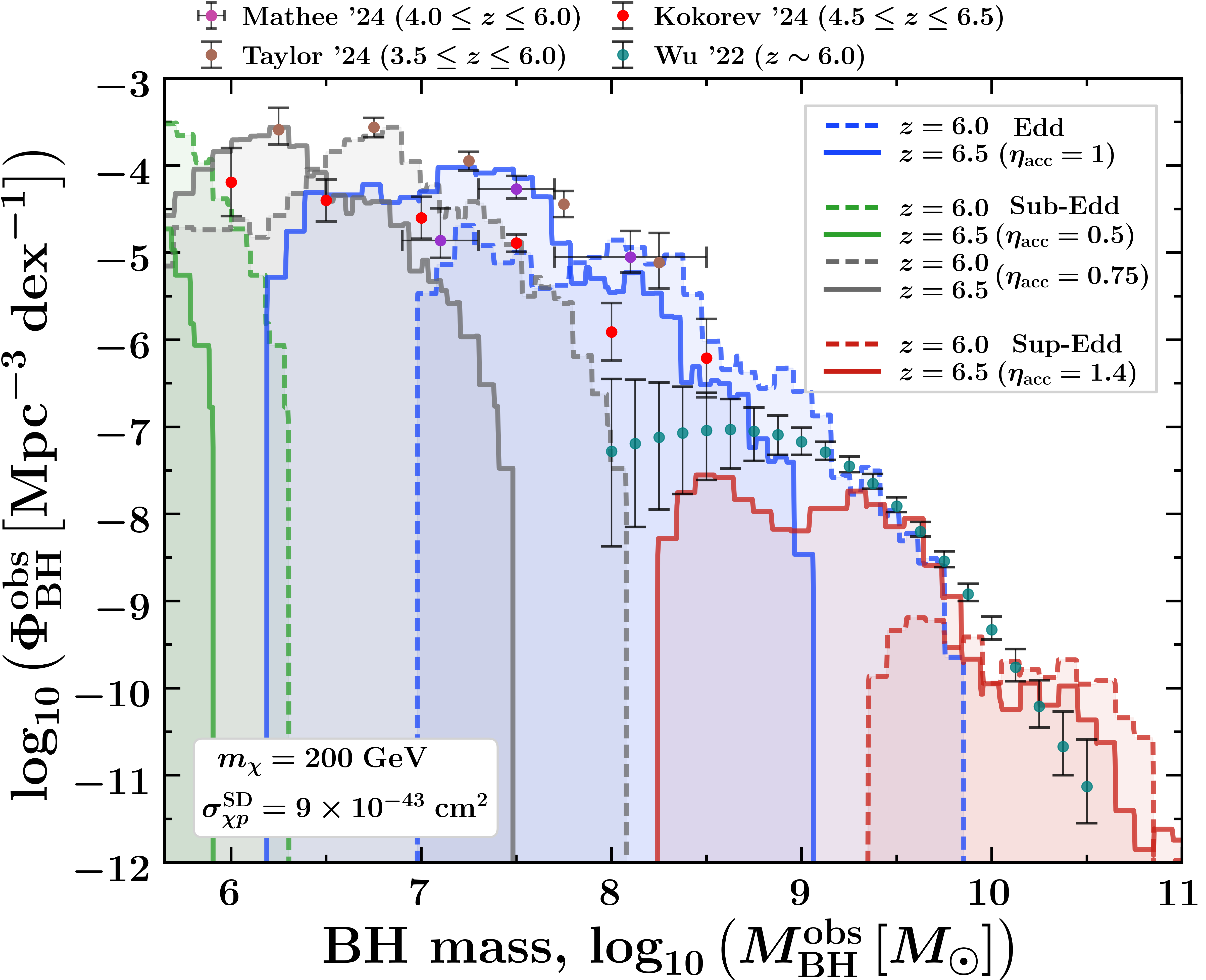}
    \caption{\justifying SMBH mass function calculated at $z_{\rm obs} = 6.0$ (dashed) and $z_{\rm obs} = 6.5$ (solid) for Eddington (blue: $\eta_{\rm acc}=1$), sub-Eddington (green: $\eta_{\rm acc}= 0.5$ and gray: $\eta_{\rm acc}= 0.75$), and super-Eddington (red: $\eta_{\rm acc}=1.4$) accretion rates, with the normalization factor ($f_{\rm norm}$) equal to $10^{-5.9}$, $10^{-5.6}$, $10^{-6.3}$, and $10^{-6.8}$, respectively. Measurements of the SMBH mass function at similar redshifts \cite{Wu:2022njo, Matthee:2023utn, Kokorev2024census, Taylor2025broad} are also displayed.}
    \label{fig:SMBH_mass_func}
\end{figure}

We derive the SMBH mass function by counting the number of SMBHs in a certain bin multiplied by the comoving number density of halos calculated with the \texttt{hmf} code\,\cite{Murray:2013qza, Murray:hmf}. In Fig.~\ref{fig:SMBH_mass_func}, we present the SMBH mass function for $z_{\rm obs}$ of $6$ and $6.5$ calculated with the benchmark parameters. We have to use a normalization factor $(f_{\rm norm} < 1)$ in order to match the observations. The $f_{\rm norm}$ can be a product of: (i) the fraction of halos that have a gNFW profile, (ii) the fraction of halos that host Pop III star at the center, and (iii) the fraction of SMBHs that are in their active phase $(f_{\rm duty})$\,\cite{Eilers:2021mdi, Pizzati:2024mwm}. In Fig.~\ref{fig:SMBH_mass_func}, we compare our predicted SMBH mass function with the observational results from\,\cite{Wu:2022njo, Matthee:2023utn, Kokorev2024census, Taylor2025broad}. These datasets also include LRDs that are believed to be associated with dust-enshrouded AGNs. We find that most of the observations of early SMBHs and LRDs can be explained by our scenario.
%
%
%%%%%%%%%%%%%%%%%%%%%%%%%%%%%%%%%%%%%%%%%%%%%%%%%%%%%%%%%%%%%%%%%%%%%%%%%%%
%%%%%%%%%%%%%%%%%%%%%%%%%%%%%%%%%%%%%%%%%%%%%%%%%%%%%%%%%%%%%%%%%%%%%%%%%%%

%%%%%%%%%%%%%%%%%%%%%%%%%%%%%%%%%%%%%%%%%%%%%%%%%%%%%%%%%%%%%%%%%%%%%%%%%%%
%%%%%%%%%%%%%%%%%%%%%       GRAVITATIONAL WAVES       %%%%%%%%%%%%%%%%%%%%%
\section{Gravitational Waves from SMBH binaries}
\label{sec:gw_signatures}
%
%
%%%%%%%%%%%%%%%%%%%%%%%%%%%%%%%%%%%%%%%%%%%%%%%%%%%%%%%%%%%%%%%%%%%%%%%%%%%
%%%%%%%%%%%%%%%%%%%%%%%%%%%%%%%%%%%%%%%%%%%%%%%%%%%%%%%%%%%%%%%%%%%%%%%%%%%
%%%%%%%%%%%%%%%%%%%%%%%%%%%%%%%%%%%%%%%%%%%%%%%%%%%%%%%%%%%%%%%%%%%%%%%%%%%
%

\begin{figure}[t]
    \centering
    \includegraphics[width=0.975\linewidth]{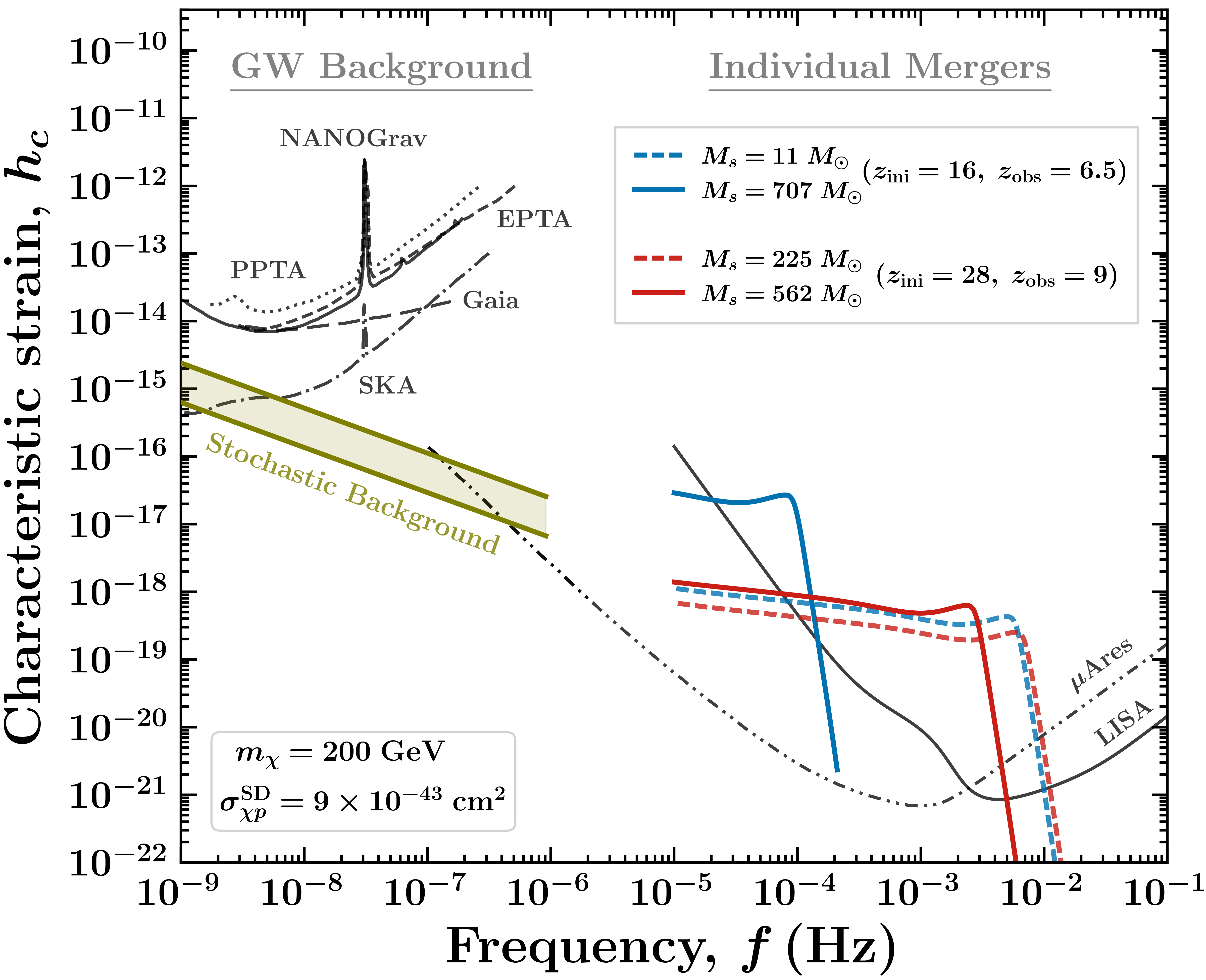}
    \caption{\justifying Characteristic strains from typical SMBH mergers are shown with solid (lowest Pop III mass) and dashed (highest Pop III mass) lines for two values of $z_{\rm in}$ and $z_{\rm obs}$ motivated from the redshift-dependent IMF in \cite{Hirano:2015wxa}. We have evolved seed BHs with Eddington accretion to form SMBHs with masses of $1.5 \times 10^6 \, M_\odot$ (blue, dashed), $9.7 \times 10^7 \, M_\odot$ (blue, solid), $1.3 \times 10^6 \, M_\odot$ (red, dashed), and $3.3 \times 10^6 \, M_\odot$ (red, solid). The stochastic GW background from Eddington-accreting SMBHs is represented by the olive shaded band. The GW detector sensitivities are illustrated with different black lines. The mass ratios for the individual mergers and stochastic background are taken to be $1$ and $1/3$, respectively.}
    \label{fig:GW_strain}
\end{figure}

Dormant SMBHs that remain electromagnetically unobserved can generate GW signals: (i) individual SMBH mergers and (ii) stochastic GW background from their mergers. In Fig.~\ref{fig:GW_strain}, we show the typical GW strain from Eddington-accreted symmetric SMBHs for the lowest and highest $z_{\rm in}$ values for an individual merger. The strain is calculated using the frequency domain IMRPhenomD\,\cite{Khan:2015jqa, Husa:2015iqa} waveform template, which is calibrated with the numerical relativity simulations, from the \texttt{pyCBC} code\,\cite{pyCBC:2024}, along with the sensitivity curves from LISA\,\cite{2017arXiv170200786A} (obtained using the fitting given in\,\cite{Robson:2018ifk, Cornish2023eXtremeGravityInstitute}), and $\mu$Ares\,\cite{Sesana:2019vho}. Similar GW signatures can also arise for other scenarios (different Pop III star mass, formation and observed redshifts, accretion rates, asymmetric mergers) within the LISA and $\mu$Ares sensitivity; however, we have not shown them in Fig.~\ref{fig:GW_strain} to maintain clarity.
%
%
%%%%%%%%%%%%%%%%%%%%%%%%%%%%%%%%%%%%%%%%%%%%%%%%%%%%%%%%%%%%%%%%%%%%%%%%%%%
%%%%%%%%%%%%%%%%%%%%%%%%%%%%%%%%%%%%%%%%%%%%%%%%%%%%%%%%%%%%%%%%%%%%%%%%%%%
%
%

To estimate the stochastic GW background from these SMBH mergers, we follow the approach outlined in\,\cite{Phinney:2001di, Jaroschewski:2022gdy}. The comoving number density of SMBH mergers is calculated using $n_{\rm SMBH}^{\rm inspiral} = n_{\rm SMBH} \times \left[ \langle \tau_{\rm insp} \rangle / \langle \tau_{\rm total} \rangle \right]$. Here, $\langle \tau_{\rm insp} \rangle$ is the average inspiral time of the SMBH binary and $\langle \tau_{\rm total} \rangle$ is the mean total time of the SMBH merger starting from halo coalescence. Their ratio is taken from Table 4 of~\cite{Jaroschewski:2022gdy} that compiles results from\,\cite{Peters:1964zz, Gergely:2007ny, Sesana:2011zv, Cavaliere:2019zzv}; the differences in the ratio values are reflected in the olive band in Fig.~\ref{fig:GW_strain}. We compute $n_{\rm SMBH}$ by evaluating the Eddington accreted SMBH mass function in the redshift range $z_{\rm obs} = 10 - 5$ in steps of $0.1$. While determining $n_{\rm SMBH}$, we include an additional factor $f_{\rm other} = f_{\rm norm}/f_{\rm duty}$, where $f_{\rm norm} = 10^{-5.9}$ (Eddington) and $f_{\rm duty}$ is taken to be $0.1$ to remain conservative \cite{Eilers:2021mdi}. This $f_{\rm other}$ includes both active and dormant SMBHs. In Fig.~\ref{fig:GW_strain}, we show the predicted stochastic GW background from these SMBHs, assuming a binary mass ratio of $1/3$, together with the $15$-year results from NANOGrav\,\cite{NANOGrav:2023ctt}. We also overlay the projected sensitivities from EPTA\,\cite{EPTA:2023fyk}, PPTA\,\cite{Reardon:2023gzh}, Gaia\,\cite{Collaboration2016TheGM}, and SKA\,\cite{Janssen:2014dka}, as compiled in~\cite{Sesana:2025udx}. Since the average masses of the sub-Eddington accreting BHs lie mostly below the SMBH mass range (see Fig.~\ref{sf:SMBH_mass_redshift}), we have not shown them. For the super-Eddington case, the GW background nearly overlaps with the Eddington band. 

The calculational details for LISA signal-to-noise ratio (SNR) and GW backgrounds are provided in the \emph{Supplemental Material}.
%
%
%%%%%%%%%%%%%%%%%%%%%%%%%%%%%%%%%%%%%%%%%%%%%%%%%%%%%%%%%%%%%%%%%%%%%%%%%%%
%%%%%%%%%%%%%%%%%%%%%%%%%%%%%%%%%%%%%%%%%%%%%%%%%%%%%%%%%%%%%%%%%%%%%%%%%%%

%
%
%%%%%%%%%%%%%%%%%%%%%%%%%%%%%%%%%%%%%%%%%%%%%%%%%%%%%%%%%%%%%%%%%%%%%%%%%%%
%%%%%%%%%%%%%%%%%%%%%%%%       CONCLUSION       %%%%%%%%%%%%%%%%%%%%%%%%%%%
\section{Conclusion}
\label{sec:conclusion}

The presence of SMBHs and LRDs (assuming an AGN origin) at very high redshifts poses a major challenge for standard astrophysical formation scenarios. While models invoking massive BH seeds can reproduce these observations, generating such seeds in the early Universe is inherently rare. We investigate the premature collapse of Pop III stars induced by the accumulation of non-annihilating DM. Focusing on the SD DM–proton scattering, we identify benchmark cross sections -- beyond current DM direct detection limits -- that induce BH formation within $\sim 10\%$ of the standard Pop III stellar lifetime.

For the benchmark DM mass and SD DM–proton scattering cross section, our framework can account for the bulk of the observed high-redshift SMBH population when allowing for Eddington, sub-Eddington, and super-Eddington accretion. We compute the SMBH mass function at $z=6$ and $6.5$, finding good agreement with current observations. In estimating the SMBH number density, we follow the evolution of host halos to infer stellar masses and compare the resulting SMBH–stellar mass relation at $z_{\rm obs} = 6$, which is consistent with the data. The framework further predicts SMBH-to-stellar mass ratios exceeding unity, especially in low-mass galaxies, providing a clear target for upcoming surveys.

Given the substantial population of SMBHs in the early Universe, our framework naturally predicts gravitational-wave signals from both individual mergers and a stochastic background. We find that the strain from individual mergers falls within the projected sensitivity of LISA, and $\mu$Ares, while the associated stochastic background is consistent with the 15-year sensitivity limits reported by the PTA collaboration, and is expected to be probed by SKA in the future. 

Taken together, these results highlight a scenario that is necessarily uncertain in its details, but rich in observational consequences. The confluence of data from JWST, DM searches, and GW observatories makes this an especially timely and promising direction to explore.

%
%
%%%%%%%%%%%%%%%%%%%%%%%%%%%%%%%%%%%%%%%%%%%%%%%%%%%%%%%%%%%%%%%%%%%%%%%%%%%
%%%%%%%%%%%%%%%%%%%%%%%%%%%%%%%%%%%%%%%%%%%%%%%%%%%%%%%%%%%%%%%%%%%%%%%%%%%

%
%
%%%%%%%%%%%%%%%%%%%%%%%%%%%%%%%%%%%%%%%%%%%%%%%%%%%%%%%%%%%%%%%%%%%
\section*{Acknowledgments}

We sincerely thank Subhadip Bouri, Prolay Chanda, Anirban Das, Raghuveer Garani, Subhajit Ghosh, Shunsaku Horiuchi, Tarak Nath Maity, Ranjini Mondol, Rohan Pramanick, Surjeet Rajendran, Nirmal Raj, Anupam Ray, Akash Kumar Saha, Tracy R. Slatyer, and Gaurav Tomar for their helpful discussions and insightful suggestions. D.B. acknowledges the Council of Scientific and Industrial Research (CSIR), Government of India, for supporting his research under the Research Associateship program through grant no.\,\,09/0079(24106)/2025-EMR-I, and support from the ISRO-IISc STC for the grant no.\,\,ISTC/PHY/RL/499. B.D. acknowledges the support of the Dept. of Atomic Energy (Govt. of India) research project RTI 4002, and the Dept. of Science and Technology (Govt. of India) for the Swarnajayanti Fellowship. R.L.\,\,acknowledges financial support from the institute start-up funds, ISRO-IISc STC for the grant no.\,\,ISTC/PHY/RL/499, and ANRF for the grant no.\,\,ANRF/ARG/2025/005140/PS.
%%%%%%%%%%%%%%%%%%%%%%%%%%%%%%%%%%%%%%%%%%%%%%%%%%%%%%%%%%%%%%%%%%%

\bibliographystyle{JHEP}
\bibliography{ref}

\clearpage
\newpage
\onecolumngrid

\begin{center}
\vspace{0.08in}

\textbf{\large{Dark Recipe for the First Giants: From Population III Stars to Early Supermassive \\[2pt] Black Holes via Dark Matter Capture}}

\vspace{0.1in}
\emph{\large{Supplemental Material}} \\[6pt]
{Sulagna Bhattacharya, Debajit Bose, Basudeb Dasgupta, Jaya Doliya, and Ranjan Laha}
\end{center}

% Reset counters and change printed numbering to include "S"
\setcounter{section}{0}
\renewcommand{\thesection}{S\arabic{section}}

\setcounter{equation}{0}
\renewcommand{\theequation}{S\arabic{equation}}

\setcounter{figure}{0}
\renewcommand{\thefigure}{S\arabic{figure}}

\setcounter{table}{0}
\renewcommand{\thetable}{S\arabic{table}}

% Numbered sections
\setcounter{secnumdepth}{3}

\section{DM Capture}
\label{sec:dm_capture}

\begin{figure}[t]
    \centering
    \begin{subfigure}{0.495\textwidth}
        \centering
        \includegraphics[width=0.975\linewidth]{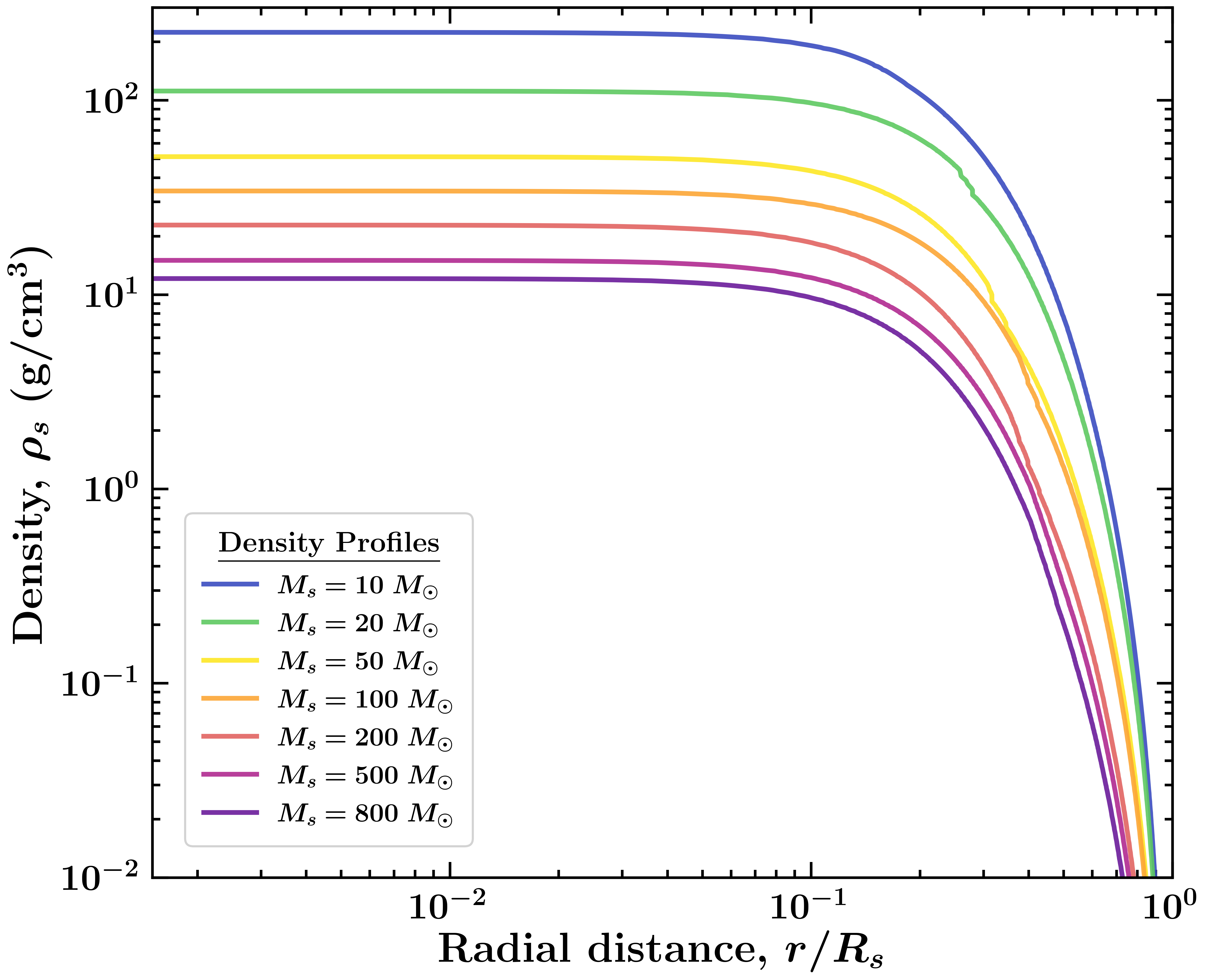}
        \caption{}
        \label{sf:pop3_density}
    \end{subfigure}
    \begin{subfigure}{0.495\textwidth}
        \centering
        \includegraphics[width=0.975\linewidth]{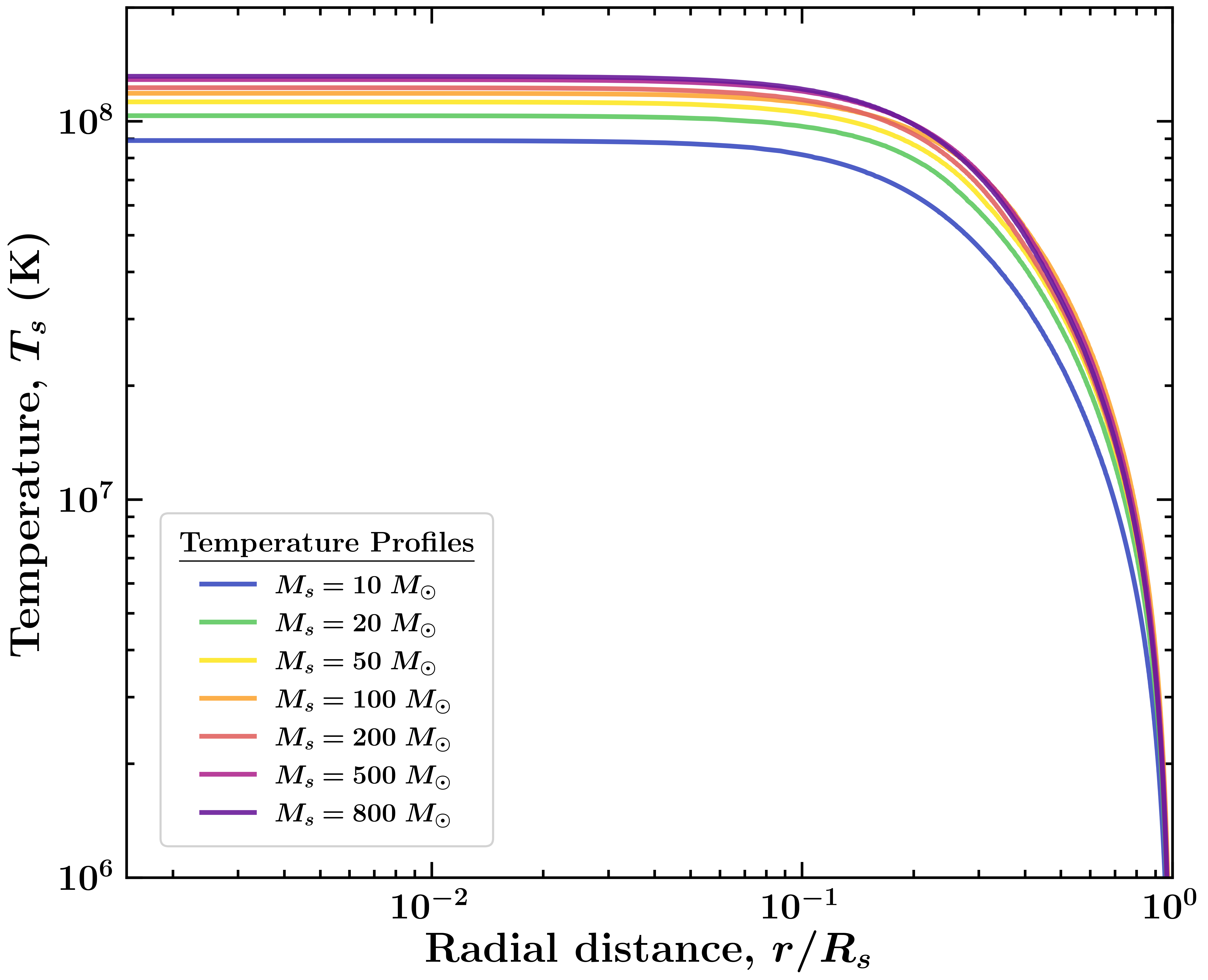}
        \caption{}
        \label{sf:pop3_temperature}
    \end{subfigure}
    \caption{\justifying The density (left) and temperature (right) profiles as a function of radial distance, obtained using \texttt{MESA}, are shown for Pop III stars with masses ranging from $10 \, M_\odot$ to $800 \, M_\odot$.}
    \label{fig:pop3_profiles}
\end{figure}

In the early Universe, DM halos of masses  $\gtrsim 10^5$–$10^6 \, {\rm M_\odot}$ are expected to harbor Pop III stars. While a star hovers through the sea of DM particles, the stellar gravitational potential focuses the DM particles toward the star. If the DM particles share non-gravitational interactions with SM states, they can scatter with stellar constituents and lose sufficient energy to become bound to the star \cite{Press:1985ug, Gould:1987ir, 1987ApJ...321..571G}. DM capture in celestial bodies can result in a broad class of observable signatures, which have been studied in\,\cite{Goldman:1989nd, Dasgupta:2019juq, Dasgupta:2020dik, Maity:2021fxw, Bose:2021yhz, Bramante:2021dyx, Bose:2021cou, Baryakhtar:2022hbu, Ray:2023auh, Bramante:2023djs, Maity:2023rez, Bose:2023yll, Bose:2024wsh, Krishna:2025ncv, Bhutani:2025jfo, Liu:2025qco, Saha:2025fgu, Bhattacharya:2025xko}. The capture rate of DM particles inside the Pop III stars is
\begin{equation}\label{eq:cap_rate}
    C_\chi = \pi R_s^2 \left( \dfrac{\rho_\chi}{m_\chi} \right) \int d^3u_\chi \dfrac{f( \vec{u}_\chi)}{u_\chi} \left( u_\chi^2 + v_{\rm esc}^2 \right) P_{\rm cap},
\end{equation}
where $R_s$ is the radius of the Pop III star, $m_\chi$ and $u_\chi$ are the mass and ambient velocity of the DM particles, respectively, and $\rho_\chi$ is the DM density in the vicinity of the Pop III star. In Eq.~\ref{eq:cap_rate}, the velocity distribution profile is taken to be a Maxwell-Boltzmann distribution $\left( f( \vec{u}_\chi ) \right)$ \footnote{Possible deviations and their effects are studied in\,\cite{Bose:2022ola}.} boosted to the rest frame of the star, $v_{\rm esc}$ is the escape velocity of the star and $P_{\rm cap} = \Theta(v_{\rm max} - u_\chi)$ is the capture probability of the DM particle with velocity $u_\chi$ and $v_{\rm max}$ is\,\cite{Ellis:2021ztw}
\begin{equation}\label{eq:v_max}
    \begin{aligned}
        & v_{\rm max} = \displaystyle\begin{dcases*}
            v_{\rm esc} \left[ \left( 1 - \dfrac{\beta_+}{2} \right)^{- n_{\rm coll}} - 1 \right]^{1/2}; m_\chi u_\chi^2 \gg m_k v_k^2 \,, \\
            v_k \, \sqrt{2 n_{\rm coll}} \left( \dfrac{m_k}{m_\chi + m_k} \right); \hspace*{0.95cm} m_\chi u_\chi^2 \ll m_k v_k^2 \,,
        \end{dcases*}
    \end{aligned}
\end{equation}
where the target particle mass is denoted by $m_k$, and the kinematic parameter $\beta_+ = 4 m_\chi m_k / (m_\chi + m_k)^2$. The velocity of the target particle ($v_k$) is obtained from the relation $\tfrac{3}{2} k_B T_s = \tfrac{1}{2} m_k v_k^2$, with $T_s$ denoting the stellar temperature, $k_B$ is the Boltzmann constant. In Eq.~\ref{eq:v_max}, $n_{\rm coll} = R_s \rho_s \sigma_{\chi k} / m_k$ is the average number of collisions a DM particle encounters while traversing the star, and $\sigma_{\chi k}$ is the scattering cross-section of DM with the target particle. In the parameter space relevant for our analysis, i.e., $m_\chi v_{\rm esc}^2 \gg m_k v_k^{2}$, the mass capture rate for $M_s = 100 \, M_\odot$ and $R_s = 6 \, R_\odot$ can be described by the approximate expression
\begin{equation}\label{eq:approx_cap_rate}
    \left. m_\chi C_\chi \right|_{\rm approx} = 6.85 \times 10^{-16} \times \left( \dfrac{\rho_\chi}{10^{11} \, {\rm GeV/cm^3}} \right) \times \left( \dfrac{200 \, {\rm GeV}}{m_\chi} \right) \times \left( \dfrac{\sigma_{\chi k}}{9.25 \times 10^{-43} \, {\rm cm^2}} \right) \ \, {\rm M_\odot/s}
\end{equation}
where the velocity dispersion is taken to be $v_d = 5 \, {\rm km/s}$. For our analysis, we compute the capture rate using Eq.~\ref{eq:cap_rate} incorporating the stellar properties of Pop III stars, and we check that for $m_\chi = 200 \, {\rm GeV}$ and $\sigma_{\chi p}^{\rm SD} = 9.25 \times 10^{-43} \ {\rm cm^2}$, the capture rates obtained from Eq.~\ref{eq:cap_rate} and \ref{eq:approx_cap_rate} differ by a factor of $3.35$. We have extracted the stellar properties by simulating the Pop III stars with Modules for Experiments in Stellar Astrophysics (\texttt{MESA})\,\cite{Paxton:2010ji, Paxton2013MODULESFE, Paxton2015MODULESFE, Paxton:2017eie, Paxton2019ModulesFE, MESA:2022zpy} with an initial metallicity of $10^{-12}$, consistent with\,\cite{Schneider2006-ge}. We track the central hydrogen mass fraction $(X_c)$ from the \texttt{history.data} file to extract details of the stellar interior. When the hydrogen fraction $X_c \lesssim 0.6$, we extract the corresponding \texttt{profile.data} file to obtain the stellar profile at that epoch. The value of $X_c$ is chosen to be $\sim 0.6$ to ensure that the Pop III stars remain within the main sequence phase of their lives. We extract the stellar radius, density, and temperature profiles from the output profile. The radial density and temperature profiles of Pop III stars are shown in Fig.~\ref{fig:pop3_profiles} for some representative stellar masses. To compute the mean density and temperature, we evaluate the average over all radial grid points provided in the output profile.
%
%
%%%%%%%%%%%%%%%%%%%%%%%%%%%%%%%%%%%%%%%%%%%%%%%%%%%%%%%%%%%%%%%%%%%%%%%%%%%
%%%%%%%%%%%%%%%%%%%%%%%%%%%%%%%%%%%%%%%%%%%%%%%%%%%%%%%%%%%%%%%%%%%%%%%%%%%
%
%
%%%%%%%%%%%%%%%%%%%%%%%%%%%%%%%%%%%%%%%%%%%%%%%%%%%%%%%%%%%%%%%%%%%%%%%%%%%%%%%%%%%%%%%%%%%%%%%
\section{Important Timescales}
%%%%%%%%%%%%%%%%%%%%%%%%%%%%%%%%%%%%%%%%%%%%%%%%%%%%%%%%%%%%%%%%%%%%%%%%%%%%%%%%%%%%%%%%%%%%%%%
%
%
\subsection{Thermalization timescale}

After being captured by Pop III stars, the DM particles initially become bound on orbits with semi-major axes larger than the stellar radius $(R_s)$, so their trajectories pass through the star. During each passage, the DM scatters with SM particles in the stellar medium and loses energy. If the velocity of DM after scattering becomes $v_{\rm final} = v_{\rm esc} \left( 1 - \xi \right)$, the total time required for the orbit to be entirely inside the star is\,\cite{Ellis:2021ztw}
\begin{equation}\label{eq:t_therm,a}
    \tau_{{\rm therm}, a} = \dfrac{\pi}{\sqrt{2} v_{\rm esc} \rho_s \sigma_{\chi k}}  \displaystyle\begin{dcases*} m_\chi \left( \dfrac{1}{ \sqrt{\xi (2-\xi)} } - 1 + \dfrac{ {\rm tanh^{-1}}(\sqrt{2}) - {\rm tanh^{-1}} \left( \dfrac{ \sqrt{2}}{ \sqrt{\xi (2-\xi)} } \right) }{\sqrt{2}} \right); m_\chi v_{\rm esc}^2 \gg m_k v_k^2 \\
    m_k \dfrac{v_{\rm esc}^2}{v_k^2} \left( \dfrac{1}{ \sqrt{\xi (2-\xi)} } - 1 \right); \hspace*{5.50cm} m_\chi v_{\rm esc}^2 \ll m_k v_k^2\,.
    \end{dcases*}
\end{equation}
After the DM orbits are confined within the stellar interior, they keep scattering with the stellar constituents and lose energy. Thus, with sufficient scatterings, the DM particles settle into the core and attain a velocity corresponding to the ambient stellar temperature, which marks the completion of the thermalization stage. The time required for the DM to drift from the outer envelope to the core of the star can be expressed as\,\cite{Ellis:2021ztw}
\begin{equation}\label{eq:t_therm,b}
    \tau_{{\rm therm},b} = \displaystyle\begin{dcases*}
        \dfrac{m_\chi \, {\rm log}\left( \dfrac{2 G m_\chi M_s}{3 R_s T_s} \right)}{4 \rho_{s,c} \sigma_{\chi k} c_s}; \hspace*{2.85cm} m_\chi \gg m_k \\
        \dfrac{m_k^2 \, {\rm log}(2)}{4 \rho_{s,c} \sigma_{\chi k} c_s m_\chi} + \dfrac{m_k \left( {\rm log}(2) + \dfrac{G M_s}{c_s^2 R_s} \right)}{\rho_{s,c} \sigma_{\chi k} c_s}; \hspace*{0.15cm} m_\chi \ll m_k
    \end{dcases*}
\end{equation}
where $\rho_{s,c}$ denotes the density of the stellar core and $c_s$ is the sound speed within the stellar medium, which is taken to be $c_s \sim v_k$. As mentioned earlier (Fig.\,\ref{fig:pop3_profiles}), the Pop III stars have been simulated using \texttt{MESA} and the corresponding stellar properties have been extracted by combining the \texttt{history.data} and \texttt{profile.data} files. As described in Eq.\,\ref{eq:t_therm,a} and \ref{eq:t_therm,b}, the total thermalization time is defined as the total time the DM particles need to form a thermalized core from the outer orbits, i.e., $\tau_{\rm therm} = \tau_{{\rm therm}, a} + \tau_{{\rm therm}, b}$.
\subsection{Collapse timescale}

For our parameter space of interest, the BH formation is triggered by the self-gravity criterion (irrespective of whether the DM is a boson or a fermion), i.e., $\rho_{\chi, c} \gg \rho_{s, c}$ with $\rho_{\chi, c}$ being the accumulated DM density at the stellar core\,\cite{McDermott:2011jp, Bell:2013xk, Dasgupta:2020mqg, Bhattacharya:2023stq, Bhattacharya:2024pmp}. Once this criterion is satisfied, the time required for the thermalized radius $r_{\rm therm}$ to contract to the Schwarzschild radius $r_{\rm Sch}$ is\,\cite{Ellis:2021ztw}
\begin{equation}\label{eq:t_coll}
    \tau_{\rm col} = \dfrac{1}{4 \rho_{s, c} c_s \sigma_{\chi k}} \displaystyle\begin{dcases*}
        m_\chi \, {\rm log} \left( \dfrac{c_s^2 r_{\rm therm}}{G M_{\rm Self}} \right); \hspace*{4.05cm} m_\chi \gg m_k \\
        m_k \left( \frac{\sqrt{2}}{c_s} - \frac{2}{c_s} \sqrt{\frac{G M_{\rm Self}}{r_{\rm therm}}} + {\rm log} \left( \frac{r_{\rm therm}}{2 G M_{\rm Self}} \right) \right); \hspace*{0.15cm} m_\chi \ll m_k
    \end{dcases*}
\end{equation}
where the thermalized radius is obtained by balancing the thermal energy and gravitational potential in the stellar core, i.e., $r_{\rm therm} = \sqrt{9 T_s / 4 \pi G \rho_{s,c} m_\chi}$.
\subsection{Bondi accretion timescale}

After the tiny BH is formed at the heart of a Pop III star, it starts accreting the stellar material and emitting Hawking radiation. As shown in\,\cite{Ellis:2021ztw}, for a typical Pop III star with a mass $\mathcal{O}(100) M_\odot$, Hawking radiation dominates only if the DM mass is close to the Planck scale. The DM is much lighter in our scenario, so Hawking radiation can be safely neglected. Although Pop III stars will continue to capture additional DM, we ignore this effect and focus solely on accretion. We model this process using Bondi accretion\,\cite{Bondi:1952ni}, which applies to non-rotating, uncharged BHs. The timescale for the tiny BH to consume the entire star and form a stellar-mass BH is
\begin{equation}\label{eq:t_Bondi}
    \tau_{\rm Bondi} = \dfrac{c_s^3}{4 \pi \alpha G^2 \rho_{s, c}} \left( \frac{1}{M_{\rm Self}} - \frac{1}{M_s} \right),
\end{equation}
where $\alpha \sim 1$\,\cite{Bondi:1952ni}, and $M_s$ and $M_{\rm Self}$ are the mass of the star and self-gravitating mass, respectively. In Table\,\ref{tab:timescale}, we show the specific timescales to form the seed BH from $100 \, M_\odot$ Pop III star, for our benchmark point ($m_{\chi}= 200$\,GeV, $\sigma_{\chi p}^{\rm SD}=9\times10^{-43}\,\rm cm^2$) used in this study. Within our parameter space of interest, the thermalization timescale dominates and governs the seed BH formation time.

\begin{table}[h]
	\centering
    \captionsetup{justification=justified, singlelinecheck=false}
	\begin{tabular}{c c}
\toprule
	\textbf{Timescales} & \hspace{0.4 cm} {in yrs} \\
\midrule
		$\tau_{\rm accum }$ & \hspace{0.4 cm} $8.8 \times 10^{-1}$ \\[0.35ex]
        $\tau_{\rm therm}$ & \hspace{0.4 cm} $1.5 \times 10^5$ \\[0.35ex]
        $\tau_{\rm col}$ & \hspace{0.4 cm} $2 \times 10^3$ \\[0.35ex]
        $\tau_{\rm Bondi}$ & \hspace{0.4 cm} $3 \times 10^{-3}$ \\[0.35ex]
        $\tau$ & \hspace{0.4 cm} $1.5 \times 10^5$ \\
\bottomrule
	\end{tabular}
	\caption{\justifying The specific timescales involved in the premature death of a $100\,M_{\odot}$ Pop III star via DM-capture induced transmutation are shown for our benchmark point ($m_{\chi}= 200$\,GeV, $\sigma_{\chi p}^{\rm SD}=9\times10^{-43}\,\rm cm^2$). The last entry of the table, $\tau$, represents the total timescale to form the seed BH as discussed in Eq.~\ref{eq:t_bhform}.}
    \label{tab:timescale}
\end{table}

%
%
%%%%%%%%%%%%%%%%%%%%%%%%%%%%%%%%%%%%%%%%%%%%%%%%%%%%%%%%%%%%%%%%%%%%%%%%%%%%%%%%%%%%%%%%%%%%%%%
%
%
\begin{figure}[t]
    \centering
    \includegraphics[width=0.475\linewidth]{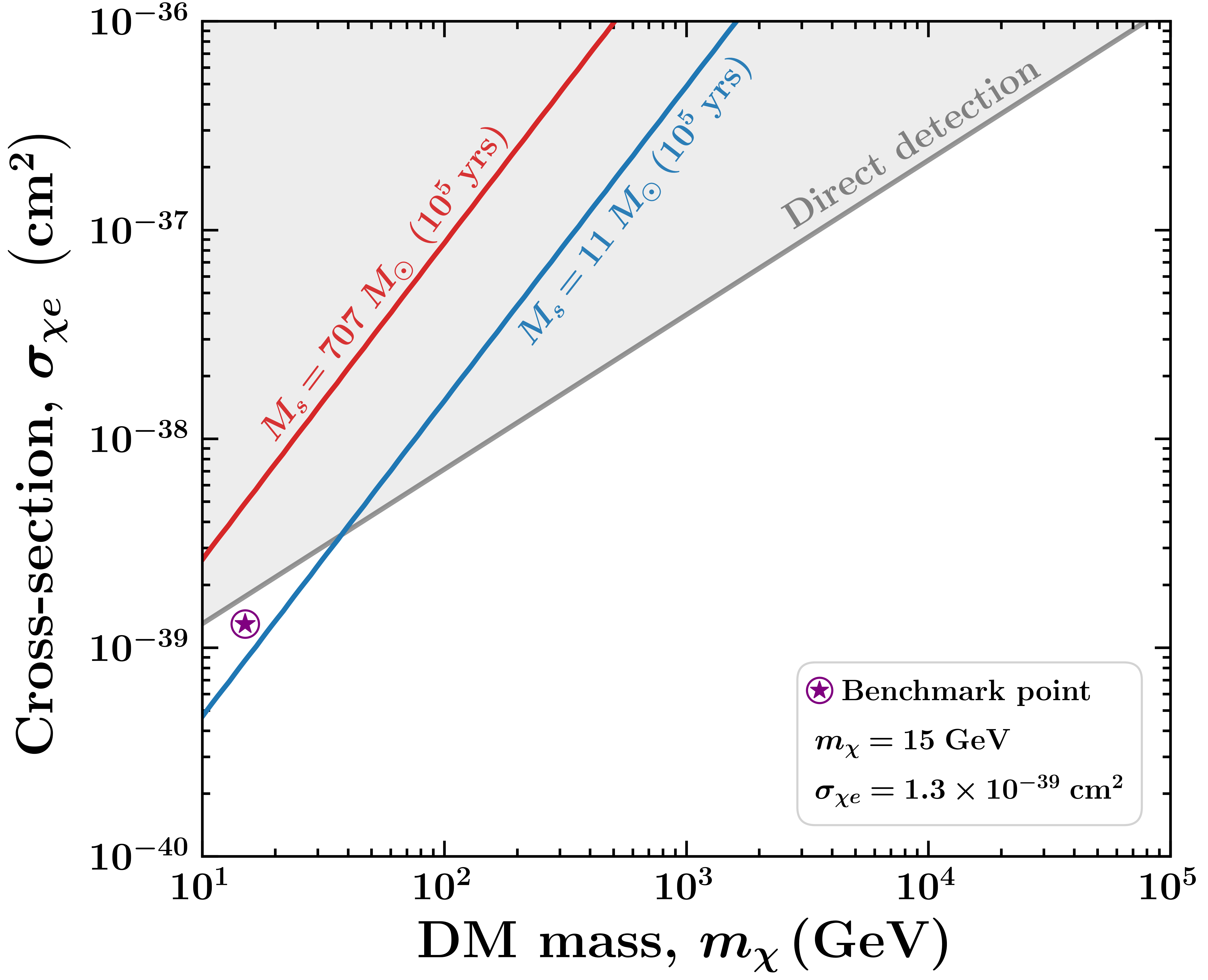}
    \caption{\justifying Contours of BH formation time of $10^5 \, {\rm yr}$ for the lowest and highest mass Pop III stars for DM-electron interaction are shown with blue and red solid lines, respectively\,\cite{Hirano:2015wxa}. The gray shaded region represents the extended direct detection limits from XENON1T\,\cite{XENON:2019gfn}. The purple point denotes the benchmark scenario that evades current constraints and successfully leads to the formation of a seed BH.}
    \label{fig:bhform_isochrone_electron}
\end{figure}
%
%
%%%%%%%%%%%%%%%%%%%%%%%%%%%%%%%%%%%%%%%%%%%%%%%%%%%%%%%%%%%%%%%%%%%%%%%%%%%%%%%%%%%%%%%%%%%%%%%
%
%
%%%%%%%%%%%%%%%%%%%%%%%%%%%%%%%%%%%%%%%%%%%%%%%%%%%%%%%%%%%%%%%%%%%%%%%%%%%%%%%%%%%%%%%%%%%%%%%
\section{Dark Matter - electron Scattering}
\label{sec:dm_electron_scattering}

We have also explored our framework for DM capture via its scattering with electrons inside the Pop III stars. To compute the electron mass density inside the stars, we use the proton number density together with the charge neutrality condition. Since the electron mass is much smaller than the proton mass, electrons are semi-relativistic in the stellar interior, and their velocities are calculated from the stellar temperature. The equal-time contours for seed BH formation (isochrones) with $\tau = 10^5 \, {\rm yrs}$ are shown in Fig.~\ref{fig:bhform_isochrone_electron}. The blue and red contours correspond to the lowest $\left(11 \, M_\odot, \, 1.5 \, R_\odot, \, \tau^{\rm Pop III}_{\rm lifetime} = 14.5 \, {\rm Myr} \right)$ and highest Pop III masses $\left(707 \, M_\odot, \, 16.4 \, R_\odot, \, \tau^{\rm Pop III}_{\rm lifetime} = 1.9 \, {\rm Myr} \right)$, respectively\,\cite{Hirano:2015wxa}. The gray shaded region denotes the direct detection limits from XENON1T\,\cite{XENON:2019gfn}, extended to higher masses using appropriate scaling. For the DM-electron benchmark point ($m_\chi = 15 \, {\rm GeV}, \sigma_{\chi e} = 1.3 \times 10^{-39} \, {\rm cm^2}$), the BH formation timescale ($\tau$) is comparable to that obtained for the benchmark parameters of DM–proton SD interaction. Consequently, all subsequent results are expected to be very similar for the case of DM–electron scattering as well. However, it should be noted that, for higher Pop III masses, the timescale would exceed $10^5 \, {\rm yrs}$ for the considered DM–electron benchmark point, which may lead to noticeable differences in other results.
%
%
%%%%%%%%%%%%%%%%%%%%%%%%%%%%%%%%%%%%%%%%%%%%%%%%%%%%%%%%%%%%%%%%%%%%%%%%%%%%%%%%%%%%%%%%%%%%%%%
\section{Dark Matter Halo}
\label{sec:dm_halo_details}
%%%%%%%%%%%%%%%%%%%%%%%%%%%%%%%%%%%%%%%%%%%%%%%%%%%%%%%%%%%%%%%%%%%%%%%%%%%%%%%%%%%%%%%%%%%%%%%
%
%

In this work, we adopt a gNFW DM halo profile, which is parametrized as\,\cite{Navarro:1995iw, Navarro:1996gj, Klypin:2000hk}
\begin{equation}\label{eq:Moore}
    \rho_\chi(r) = \dfrac{\rho_0}{ \left( \dfrac{r}{r_s} \right)^\gamma \left[ 1 + \left( \dfrac{r}{r_s} \right) \right]^{3 - \gamma} },
\end{equation}
where $\rho_0$, $r_s$ are the characteristic density and the scale radius, respectively, and the inner slope is fixed at $\gamma = 3/2$\,\cite{Moore:1999gc}. However, we have checked that for $\gamma \geq 1.24$, the required density $\rho_\chi = 10^{11} \, {\rm GeV/cm^3}$ can be reached at distances exceeding $10$ times the typical radius of a Pop III star, i.e., $50 \, R_\odot$. Therefore, our analysis remains valid for gNFW DM profiles with $\gamma \geq 1.24$. We rewrite the density profile so that it is expressed directly in terms of the halo mass $(M_{200})$ and the concentration parameter $(c_{200})$. The parameters $\rho_0$ and $r_s$ are then determined using two conditions: (i) the mean density within $r_{200}$ satisfies $\bar{\rho}(r_{200}) = 200 \, \rho_{\rm crit}$, where $\rho_{\rm crit}$ is the critical density of the universe at that epoch and $r_{200} = c_{200} \, r_s$, and (ii) the enclosed mass within $r_{200}$ is $M_{200} = \int_0^{r_{200}} 4 \pi r^2 \, \rho_\chi(r) \, dr$. Using this density profile, we determine the radial position in the halo where $\rho_{\chi} = 10^{11} \, {\rm GeV}$, i.e., $r_{11}$, and assume that the Pop III star is located at this position. Our framework is insensitive to the precise location of the Pop III star, and remains valid for any position within the radius $r_{11}$. We need to estimate the typical halo velocities to calculate the capture rate at that location in the halo. The halo escape velocity at a distance $r$ from the center is $u_{\rm esc} = \sqrt{- 2 V(r)}$, where $V(r)$ is the gravitational potential. We have estimated the halo rotational velocity at the same location using the relation\,\cite{vandenBosch:1999ka}
\begin{equation}\label{eq:rot_velocity}
    u_{\rm rot}(r) = \sqrt{\dfrac{r_{200} \ \, \upsilon  \left( \dfrac{r \, c_{200}}{r_{200}} \right)}{r \ \, \upsilon(c_{200})}}, \hspace{0.25cm} {\rm with \ \ } \upsilon(y) = \int_0^y x^{2 - \gamma} (1+y)^{\gamma - 3} \, dy.
\end{equation}
Among the relevant timescales, only the accumulation timescale $(\tau_{\rm accum})$ depends on the DM density profile. As shown in Table~\ref{tab:timescale}, $\tau_{\rm accum}$ is much smaller than the thermalization timescale ($\tau_{\rm therm}$). We have checked that the total timescale for the benchmark point ($m_{\chi}= 200$\,GeV, $\sigma_{\chi p}^{\rm SD}=9\times10^{-43}\,\rm cm^2$) remains approximately unchanged even when the Pop III star is placed at a location $(r_9, \ {\rm i.e., where} \ \rho_\chi = 10^9 \, {\rm GeV/cm^3}$), for which the condition $r_9 = 50 \times R_\odot$ can be achieved even for the NFW profile, i.e., $\gamma = 1$.

In\,\cite{Fakhouri:2010st}, a fitting formula for the halo growth rate as a function of the halo mass and the redshift is provided, considering growth via both accretion and mergers. To model the growth of our halos, we adopt the mean fitting formula expressed as\,\cite{Fakhouri:2010st}
\begin{equation}\label{eq:halo_growth}
    \left\langle \dfrac{dM_{\rm halo}}{dt} \right\rangle = 46.1 \, \left( \dfrac{M_{\rm halo}}{10^{12} \, {\rm M_\odot}} \right)^{1.1} \ (1 + 1.11 \, z) \ \sqrt{\Omega_m (1+z)^3 + \Omega_\Lambda} \ \  {\rm M_\odot /yr}.
\end{equation}
Since the fitting relation is shown only up to $z = 14$ in\,\cite{Fakhouri:2010st}, for redshifts beyond this limit, we keep the halo accretion rate fixed at its maximum value at $z = 14$ instead of allowing it to increase monotonically at higher redshifts.
%
%
%%%%%%%%%%%%%%%%%%%%%%%%%%%%%%%%%%%%%%%%%%%%%%%%%%%%%%%%%%%%%%%%%%%%%%%%%%%

%
%
%%%%%%%%%%%%%%%%%%%%%%%%%%%%%%%%%%%%%%%%%%%%%%%%%%%%%%%%%%%%%%%%%%%%%%%%%%%%%%%%%%%%%%%%%%%%%%%
\section{Detailed Explanation of Figure \ref{fig:SMBH_mass_redshift_MBH_Mstar_combined}}
\label{sec: Figures explanation}
%%%%%%%%%%%%%%%%%%%%%%%%%%%%%%%%%%%%%%%%%%%%%%%%%%%%%%%%%%%%%%%%%%%%%%%%%%%%%%%%%%%%%%%%%%%%%%%
%
%
\subsection{Black Hole Mass vs Observed Redshift}

Fig.~\ref{sf:SMBH_mass_redshift} represents the observed mass of the BH, formed through DM capture inside Pop III stars, as a function of $z_{\rm obs}$. Different bands correspond to different accretion rates, $\eta_{\rm acc}$, for the evolution of seed BHs. Each band is an envelope of sub-bands that correspond to four different $z_{\rm in}$ values (16, 20, 24, 28). The lower and upper edges of these sub-bands correspond to the lowest and highest mass Pop III stars, respectively, at that $z_{\rm in}$ (see Fig.~17 of\,\cite{Hirano:2015wxa}). Therefore, the lower and upper edges of the bands in Fig.~\ref{sf:SMBH_mass_redshift} represent the lowest mass Pop III star (at $z_{\rm in} = 16$) and the highest mass Pop III star (at $z_{\rm in} = 28$), respectively. We also plot the observational data and find that most of the observations can be explained by this mechanism.
%
%
%%%%%%%%%%%%%%%%%%%%%%
%
%s
\subsection{Black Hole Mass vs Stellar Mass}

Fig.~\ref{sf:MBH_vs_Mstar} shows the relation between the BH mass and the host galaxy stellar mass at $z_{\rm obs} = 6$ for four different seed BH accretion efficiencies, indicated by different colors. We consider halo masses in the range $M_{\rm halo} \in [10^6, 10^9] \, M_{\odot}$. After tracking the halo growth up to $z_{\rm obs}$ with Eq.~\eqref{eq:halo_growth}, we compute the galaxy stellar mass using the fitting relation from \texttt{Data Release\,1} of the \texttt{UniverseMachine} code\,\cite{Behroozi2019_UM_DR1}. Each color contains several sub-bands determined by the initial redshifts of the Pop III stars. For each sub-band, the horizontal width arises from changes in the halo mass, while the vertical width reflects differences in the initial Pop III star mass, taken from \,\cite{Hirano:2015wxa}. We impose an upper cut on the BH mass of $10^{11} \, M_{\odot}$ and on the host galaxy stellar mass of $10^{12} \, M_{\odot}$. In addition, we apply the condition $M_{\rm BH}^{\rm obs} \leq M_{\rm halo}^{\rm obs}$, which deforms the shape of some sub-bands away from being rectangular.
%
%
%%%%%%%%%%%%%%%%%%%%%%%%%%%%%%%%%%%%%%%%%%%%%%%%%%%%%%%%%%%%%%%%%%%%%%%%%
%
%
%%%%%%%%%%%%%%%%%%%%%%%%%%%%%%%%%%%%%%%%%%%%%%%%%%%%%%%%%%%%%%%%%%%%%%%%%%%%%%%%%%%%%%%%%%%%%%%
\section{Gravitational Waves}
%%%%%%%%%%%%%%%%%%%%%%%%%%%%%%%%%%%%%%%%%%%%%%%%%%%%%%%%%%%%%%%%%%%%%%%%%%%%%%%%%%%%%%%%%%%%%%%
%
%
%%%%%%%%%%%%%%%%%%%%%%%%%%%%%%%%%%%%%%%%%%%%%%%%%%%%%%%%%%%%%%%%%%%%%%%%%%%
\begin{figure}[t]
    \centering
    \begin{subfigure}{0.495\textwidth}
        \centering
        \includegraphics[width=0.975\linewidth]{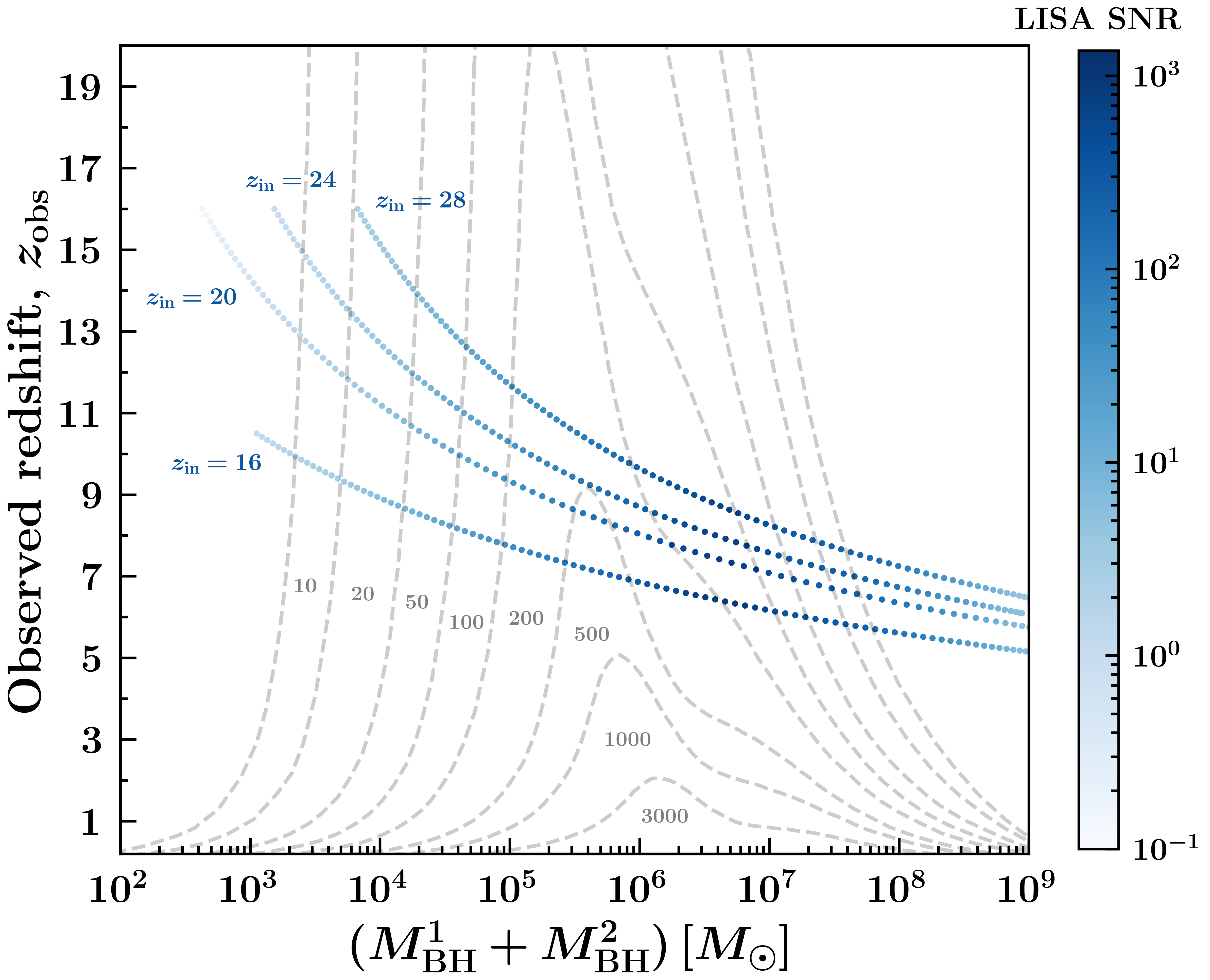}
        \caption{}
        \label{sf:LISA_SNR_low}
    \end{subfigure}
    \begin{subfigure}{0.495\textwidth}
        \centering
        \includegraphics[width=0.975\linewidth]{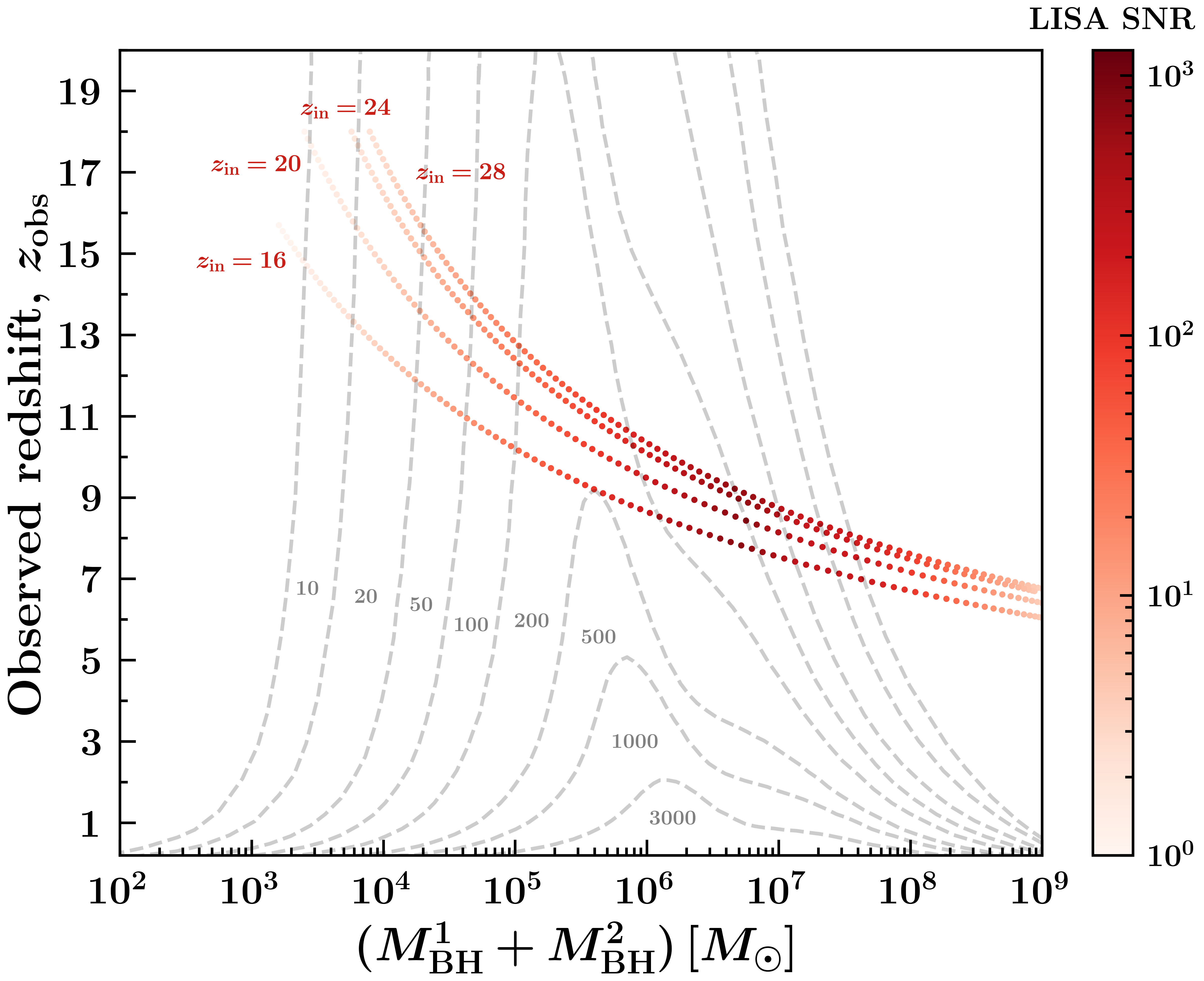}
        \caption{}
        \label{sf:LISA_SNR_high}
    \end{subfigure}
    \caption{\justifying Color plot of the LISA SNR in the observed redshift and total BH mass plane. The color bars correspond to the SNR values of the lowest (left) and highest (right) Pop III star masses. For each mass case, four distinct tracks represent different initial redshifts $(z_{\rm in} =16, \, 20, \, 24, \,28)$. The gray dashed lines indicate contours of constant LISA SNR\,\cite{2017arXiv170200786A}.}
    \label{fig:LISA_SNR}
\end{figure}
%%%%%%%%%%%%%%%%%%%%%%%%%%%%%%%%%%%%%%%%%%%%%%%%%%%%%%%%%%%%%%%%%%%%%%%%%%%
%
%

In this section, we will outline the calculation of GW signatures of SMBH mergers. For individual mergers, we compute the gravitational wave amplitudes, $\tilde{h}(f)$, with the \texttt{pyCBC} code\,\cite{pyCBC:2024} and the corresponding sky-averaged LISA SNR is calculated by the formula\,\cite{Robson:2018ifk}
\begin{equation}
    {\rm SNR} = \dfrac{16}{5} \int_0^\infty \, df \, \dfrac{\left| \tilde{h}(f) \, df \right|^2}{S_n(f)},
\end{equation}
where $S_n(f)$ denotes the one-sided power spectral density of detector noise (for LISA) taken from\,\cite{Robson:2018ifk}. In Fig.~\ref{fig:GW_strain}, we present the characteristic strain for representative mergers expected in our scenario for the lowest and highest $z_{\rm in}$. To compute the SNR, we use all four benchmark $z_{\rm in}$ values, i.e., (16, 20, 24, 28), mentioned in the manuscript. After the seed BH forms from a Pop III star with the benchmark SD DM-proton scattering cross-section, we choose the subsequent Eddington accretion of the seed to calculate the SNR of symmetric individual mergers. We scan over the observed redshifts and calculate $\tilde{h(f)}$ using \texttt{pyCBC}, incorporating the luminosity distance, to check whether the signal amplitude falls within the LISA sensitivity window. We then compute the SNR for those mergers that have at least partial overlap with the LISA band. In Fig.~\ref{fig:LISA_SNR}, we show the resulting SNR across the scanned redshifts, considering both the lightest (blue) and the heaviest (red) Pop III star masses in the left and right panels, respectively.

We estimate the GW background using the framework of\,\cite{Jaroschewski:2022gdy}, which is based on\,\cite{Phinney:2001di}. The characteristic strain of the GW background is given by\,\cite{Phinney:2001di, Jaroschewski:2022gdy}
\begin{equation}\label{eq:gw_back}
    h_{c, {\rm back}} = 3 \times 10^{-24} \left( \dfrac{\mathcal{M}_c}{M_\odot} \right)^{5/6} \left( \dfrac{f}{1 \ {\rm mHz}} \right)^{-2/3} \left( \dfrac{n_{\rm SMBH}^{\rm merge}}{\rm Mpc^{-3}} \right)^{1/2} \left( \dfrac{\left\langle (1+z)^{-1/3} \right\rangle}{0.74} \right)^{1/2},
\end{equation}
where $\mathcal{M}_c$ denotes the chirp mass. For two SMBHs merging with masses $M_{\rm BH}^1$ and $M_2 = \lambda M_{\rm BH} ^2$, the resulting SMBH mass after the merger can be written as\,\cite{Jaroschewski:2022gdy}
\begin{equation}\label{eq:MBH_final}
    M_{\rm final} = \left[ 1 + \zeta_{\rm acc} + 0.1 \, \kappa(\lambda) \right] (1 + \lambda) \, M_{\rm BH}^1,
\end{equation}
where $\kappa = \sqrt{\lambda^3/(1+\lambda)^6}$ and $\zeta_{\rm acc}$ is the fraction of mass accreted by the SMBHs during the merger, and we have fixed it at a constant value of $0.33$\,\cite{Jaroschewski:2022gdy}. Thus, the chirp mass in Eq.~\eqref{eq:gw_back} is expressed as
\begin{equation}\label{eq:M_chirp}
    \mathcal{M}_c = \dfrac{\kappa(\lambda)}{1 - 0.1 \, \kappa(\lambda)} \times M_{\rm final}.
\end{equation}
In Eq.~\eqref{eq:gw_back}, the redshift average $\left\langle (1+z)^{-1/3} \right\rangle$ is taken to be $0.8$ that is valid for a flat universe. As shown in\,\cite{Phinney:2001di}, this value depends only mildly on the number density distribution of the mergers. For the SMBH merger mass ratios, we consider $\lambda$ to be equal to $1/3$ and compute the GW background as described in\,\cite{Jaroschewski:2022gdy}. We find that for $\lambda$ equal to $1/3$ and $1/30$, the corresponding strain amplitudes lie in nearly the same range, and given the uncertainties in the merger timescales, the resulting bands largely overlap. Therefore, we show only the strain for $\lambda = 1/3$ in Fig.~\ref{fig:GW_strain}.
%
%
%%%%%%%%%%%%%%%%%%%%%%%%%%%%%%%%%%%%%%%%%%%%%%%%%%%%%%%%%%%%%%%%%%%%%%%%%
%
%

\end{document}